\documentclass[onecolumn]{emulateapj}

\newcommand{\cm}{{\rm\,cm}}

\newcommand{\second}{{\rm\,s}}
\newcommand{\kms}{{\rm\,km\,s^{-1}}}

\newcommand{\au}{{\rm\,AU}}
\newcommand{\pc}{{\rm\,pc}}

\newcommand{\msun}{{\rm\,M_\odot}}

\newcommand{\yr}{{\rm\,yr}}
\newcommand{\gm}{{\rm\,g}}

\newcommand{\gauss}{{\,\rm G}}
\newcommand{\muG}{{\,\mu\rm G}}

\newcommand{\ct}{\citealt}

\begin{document}

\title{Protostellar Accretion Flows Destabilized by Magnetic Flux
  Redistribution}

\author{Ruben Krasnopolsky\altaffilmark{1}, Zhi-Yun Li\altaffilmark{2}, Hsien Shang\altaffilmark{1}, Bo Zhao\altaffilmark{2}}
\altaffiltext{1}{Academia Sinica, Institute of Astronomy and Astrophysics, Taipei, Taiwan}
\altaffiltext{2}{University of Virginia, Astronomy Department,
  Charlottesville, VA, USA}

\shortauthors{{\sc Krasnopolsky et al.}}
\shorttitle{{\sc Unstable Magnetized Protostellar Accretion}}

\begin{abstract}

Magnetic flux redistribution lies at the heart of the problem of star
formation in dense cores of molecular clouds that are magnetized to a
realistic level. If all of the magnetic flux of a typical core were
to be dragged into the central star, the stellar field strength would
be orders of magnitude higher than the observed values. This
well-known ``magnetic
flux problem'' can in principle be resolved through non-ideal MHD
effects. Two dimensional
(axisymmetric) calculations have shown that ambipolar diffusion,
in particular,
can transport magnetic flux outward relative to matter, allowing
material to enter the central object without dragging the field
lines along. We show through simulations that such axisymmetric
protostellar accretion flows are unstable in three
dimensions to magnetic interchange instability in the
azimuthal direction. The instability is driven by the magnetic
flux redistributed from the matter that enters the central object.
It typically starts to develop during the transition from the
prestellar phase of star formation to the protostellar mass
accretion phase. In the latter phase,
the magnetic flux is transported outward mainly through advection,
by strongly magnetized low-density regions that expand against
the collapsing inflow. The tussle between the gravity-driven
infall and magnetically driven expansion leads to a highly
filamentary inner accretion flow that is more disordered than
previously envisioned. The efficient outward transport of magnetic
flux by advection lowers the field strength at small radii, making
the magnetic braking less efficient and the formation of rotationally
supported disks easier in principle. However, we find no evidence
for such disks in any of our rotating collapse simulations. We
conclude that the inner protostellar accretion flow is shaped
to a large extent by the flux redistribution-driven magnetic
interchange instability. How disks form in such an environment
is unclear.
\end{abstract}


\section{Introduction}
\label{intro}

Star-forming dense cores of nearby molecular clouds are observed to
be significantly magnetized. One line of evidence is the
polarization
of their submillimeter dust emission (\ct{WardThompson+2000}), which
indicates the existence of an ordered magnetic field on the
$0.1\pc$ scale (e.g., \ct{Matthews+2009}; \ct{Davidson+2011}) and
smaller (e.g., \ct{Girart+2006}). Another line of evidence comes
from Zeeman measurement. \citet{TrolandCrutcher2008} carried out the
most extensive OH Zeeman survey of the dark cloud cores to date with
the Arecibo telescope. The measured line-of-sight field strengths
(within a beam size of $3\arcmin$) lie between $\sim 10$ to
$\sim 25 \muG$. The inferred mean mass-to-flux ratio is
$\lambda_{\rm{los}} \sim 4.8$ (in units of the critical
mass-to-flux ratio $1/[2\pi G^{1/2}]$, \ct{NakanoNakamura1978}),
based on the measured line-of-sight field strength and column
density. Geometric corrections would bring the ratio closer to the
critical value, by a typical factor of 2--3 (\ct{Shu+2000};
\ct{TrolandCrutcher2008}). Dense cores therefore appear to be moderately
strongly magnetized, with an intrinsic dimensionless mass-to-flux
ratio $\lambda$ of a few to several. Such a magnetic field is too
weak to prevent the core from forming one or more stars through
gravitational collapse. It is, however, strong enough to affect, even
control, the dynamics of the core collapse and protostellar mass
accretion, especially the inner part of the protostellar accretion
flow that is directly relevant to disk formation.

How rotationally supported disks form around protostars is still
uncertain. If an ordered magnetic field of the observationally inferred
strength is strictly frozen into the collapsing core material
(i.e., in the ideal MHD limit), it would completely suppress the
formation of a rotationally supported disk through excessive
magnetic braking (\ct{Allen+2003}; \ct{Galli+2006}; \ct{MellonLi2008};
\ct{HennebelleFromang2008}; \ct{Seifried+2011}; \ct{Dapp+2012}; see,
however, \ct{Machida+2011}, \ct{Duffin+2011} and \ct{Seifried+2012}
for a different view). The reason is that the matter that enters
the central object would drag its frozen-in magnetic field into a
split-monopole, which is strong enough close to the protostar to
brake the rotation of the protostellar accretion flow completely
(\ct{Galli+2006}). Flux-freezing would also lead to the well-known
``magnetic flux problem'' in star formation, namely, if the
magnetic flux of a typical star-forming core is carried into
the central star, the stellar field strength would be
orders of magnitude higher than the observed values (e.g.,
\ct{Nakano1984}, see his \S 4). To resolve both problems, the
field lines must be allowed to move relative to the bulk matter,
i.e., the magnetic flux must be redistributed.

In lightly-ionized dense cores, magnetic flux redistribution can be
achieved through non-ideal MHD effects,
including ambipolar diffusion, Ohmic dissipation and the Hall effect
(e.g., \ct{Nakano+2002}; \ct{KunzMouschovias2010}). Axisymmetric
(1D and 2D) calculations have shown that ambipolar diffusion, in
particular, can enable the core material to collapse onto the
central stellar object without dragging the field lines along
(\ct{CiolekKonigl1998};
 \ct{KrasnopolskyKonigl2002}; \ct{MellonLi2009}; \ct{Li+2011};
\ct{BraidingWardle2012}). The redistributed stellar magnetic flux
is trapped instead by the ram pressure of the collapsing flow
(\ct{LiMcKee1996}), creating a circumstellar region where the
flow dynamics is magnetically controlled. In this region, the
rotation is nearly completely braked for a realistic level of
initial core magnetization (making the formation of rotationally
supported disks difficult; \ct{KrasnopolskyKonigl2002};
\ct{MellonLi2009}; \ct{Li+2011}) and the material in the dense
equatorial region becomes magnetically supported (reducing
the infall speed well below the local free-fall value). The
conclusion from the axisymmetric calculations is that the
ambipolar diffusion-enabled redistribution of the magnetic flux
that would have entered the protostar in the ideal MHD limit
makes the magnetic field dynamically more important outside the
central object compared to the ideal MHD case. Similar results
were found for Ohmic dissipation and the Hall effect (e.g.,
\ct{Li+2011}).

An important issue that has not been fully addressed is the
stability of the circumstellar structure produced by the
magnetic flux redistribution under the assumption of
axisymmetry. It has been suspected for some time that the
structure may be prone to the magnetic interchange instability
once the axisymmetry is removed, because the region close to
the protostar is expected to be more strongly magnetized
than farther out (\ct{LiMcKee1996}, see their Fig.\ 1;
\ct{CiolekKonigl1998}; \ct{KrasnopolskyKonigl2002}).
The criterion for the interchange instability is that the mass-to-flux
ratio decreases in the direction of gravity in the simplest case of
a magnetically supported, non-rotating sheet (\ct{SpruitTaam1990}).
This criterion is formally satisfied in part of the 1D (axisymmetric)
ambipolar diffusion-mediated accretion flow studied by
\citet{CiolekKonigl1998}. However, the development of the expected
instability has never been explored in detail using 3D non-ideal
MHD simulations. It is the goal of this paper.

Our non-ideal MHD simulations will build on the work of \citet{Zhao+2011},
who investigated the collapse of magnetized cores and
protostellar accretion using an ENZO-based ideal MHD code
(\ct{WangAbel2009}). The magnetic flux redistribution is
achieved through a sink particle
treatment. When the mass in a cell is accreted onto a sink particle,
the magnetic field is left behind in the cell (see also
\ct{Seifried+2011}); the treatment is a crude
representation of the field-matter decoupling expected at high
densities (of order $10^{12}\cm^{-3}$ or higher; \ct{Nakano+2002};
\ct{KunzMouschovias2010}). The decoupled magnetic flux piles up
near the sink particle, leading to a high magnetic pressure
that is released through the escape of field lines along the
directions of least resistance. The net result is that the magnetic
flux dragged into the decoupling region near the protostar
along some azimuthal directions by the collapsing flow is advected
back out along other directions in highly magnetized, low-density,
expanding regions. The main effects of the magnetic flux
redistribution on the protostellar accretion flow are (1) the
co-existence of the magnetically driven expansion and
gravitationally driven infall, which makes the flow more disordered
than previously envisioned, and (2) advective transport of the
redistributed magnetic flux to large distances, which is absent
under the assumption of axisymmetry. We show in this paper that
these two basic features are preserved in the presence of the
two most widely studied non-ideal MHD effects in star formation:
ambipolar diffusion and Ohmic dissipation.

\section{Problem Setup}
\label{setup}

Following \citeauthor{Krasnopolsky+2010} \citeyearpar{Krasnopolsky+2010,
Krasnopolsky+2011} and \citet{Li+2011},
we start our simulations from a uniform, spherical
core of $1\msun$ and radius $10^{17}\cm$. The initial density is
therefore $\rho_0=4.77 \times 10^{-19}\gm\cm^{-3}$, corresponding
to a molecular hydrogen number density of $10^5\cm^{-3}$. We adopt
an isothermal equation of state, with a sound speed $a=0.2\kms$,
up to a critical density $\rho_c=10^{-13}\gm\cm^{-3}$. Beyond
$\rho_c$, a polytropic equation of state $p\propto\rho^{5/3}$ is adopted.
At the beginning of the simulation, we impose a uniform magnetic field
of $B_0=35.4\muG$. It corresponds to a dimensionless mass-to-flux
ratio of $\lambda=2.92$ for the core as a whole (and a plasma $\beta$
of 3.82 for the adopted isothermal sound speed), which is in the
observationally inferred range
(see \S \ref{intro}). The mass-to-flux ratio for the central flux tube is
$4.38$, higher than the global value by $50\%$. We have experimented
with magnetic fields as weak as $3.54\muG$, and found qualitatively
similar results.

For illustrative purposes, we adopt the simplified treatment of
ambipolar diffusion of \citeauthor{Shu1992} (\citeyear{Shu1992}, Chapter 27),
with the magnetic field tied to the ions and the ion density
proportional to the square root of the mass density. The
proportionality constant scales with $\zeta^{1/2}$,
where $\zeta$ is the cosmic ray ionization rate. We will consider
the canonical value $\zeta=10^{-17}\second^{-1}$, although there is
evidence for higher values in star-forming clouds (e.g.,
\ct{Padovani+2009}). Our reference model will have $\zeta=9\times
10^{-17}\second^{-1}$, which contains three times more ions than in
the canonical case. In addition, we will consider cases with a
spatially constant resistivity $\eta=10^{17}\cm^2\second^{-1}$. The value
is larger than the classical microscopic resistivity at the
densities encountered in our simulations (see \ct{Li+2011}). It
is chosen to illustrate the effects of Ohmic dissipation while
minimizing the violent numerical reconnection that dominates
the protostellar accretion simulations in the ideal MHD limit
(e.g., \ct{MellonLi2008}). We also consider cases where
a relatively large resistivity of $\eta=10^{19}\cm^2\second^{-1}$
is assumed within a small radius of $2\times 10^{14}\cm$ of the
central object, to illustrate the effects of magnetic decoupling
(see \S \ref{step}). In some cases, we include an initial solid-body rotation
of angular speed $\Omega_0=10^{-13}\second^{-1}$ in the core, to study
the possibility of disk formation.
It corresponds to a ratio of rotational to gravitational binding energy
of 0.025, which is typical of the values inferred for NH$_3$ cores
(\ct{Goodman+1993}). The models to be discussed in the result sections
(\S \ref{stability} and \S \ref{filamentary}) are listed in Table \ref{table:first}.

\begin{deluxetable}{lllll}
\tablecolumns{6}
\tablecaption{Parameters of 3D Non-Ideal MHD Models \label{table:first}}
\tablehead{
\colhead{Model}
& \colhead{$\zeta$ ($10^{-17}\second^{-1}$)}
& \colhead{$\eta$ (${\rm{cm}}^2\second^{-1}$)}
& \colhead{$\Omega_0$ ($10^{-13}\second^{-1}$)}
& \colhead{Central object$^a$ }
}
\startdata
A   & 9  & 0  & 0  &  after   \\
B   & 9  & 0  & 0  &  before  \\
C   & 9  & 0  & 1  &  before  \\
D   & 1  & 0  & 0  &  before  \\
E   & no AD  & spatially uniform, $10^{17}$  & 0  &  before  \\
F   & no AD  & spatially uniform, $10^{17}$  & 1  &  before  \\
G   & no AD  & step function, 1 \& $10^{19}$ & 0  &  before  \\
H   & no AD  & step function, 1 \& $10^{19}$ & 1  &  before  \\
I   & 9  & step function, 1 \& $10^{19}$ & 0  &  before  \\
J   & 9  & step function, 1 \& $10^{19}$ & 1  &  before  \\
\enddata
\tablecomments{a). The 3D simulations are restarted from their
corresponding 2D (axisymmetric) simulations either before or
after the formation of an object of significant mass at
the center.}
\end{deluxetable}

Our 3D non-ideal MHD simulations were carried out in the coordinate
system most natural for the collapse problem: the spherical polar
system $(r, \theta, \phi)$. Both the initial magnetic field and rotation
directions are along the $\theta=0$ axis. We choose a non-uniform grid of
$96\times 64\times 60$. In the radial direction, the inner and
outer boundaries are located at
$r=10^{14}$ and 10$^{17}\cm$, respectively. The radial cell size
is smallest near the inner boundary ($5\times 10^{12}\cm$ or
$\sim 0.3\au$). It increases outward by a factor of $1.08$
between adjacent cells. In the polar direction, we choose a
relatively large cell size ($7.5^\circ$) near the polar axes,
to prevent the azimuthal cell size from becoming prohibitively
small, because the time step must be proportional to the cell
size squared to ensure numerical stability for our explicit
treatment of the non-ideal MHD effects, particularly ambipolar
diffusion. The polar cell size decreases smoothly to a minimum of
$0.63^\circ$ near the equator, where most of the protostellar
mass is accreted, through the ``pseudo-disk'' (\ct{GalliShu1993}).
The grid is uniform in the azimuthal direction,
with the cell size equal to 6 degrees. For our
reference model (Model B in Table \ref{table:first}), we have increased the
number of azimuthal cells to 90 and 120, and found qualitatively
similar results.

The boundary conditions in the azimuthal direction are periodic.
In the radial direction, we impose the standard outflow
boundary conditions. Material leaving the inner radial boundary is
collected as a point mass (protostar) at the center. It acts on
the matter in the computational domain through gravity. On the
polar axes, the boundary condition
is chosen to be reflective (as in the 2D axisymmetric case).
Although this is not strictly valid in 3D, we do not expect
it to affect much the dynamics of the equatorial region, through
which most of the mass accretion onto the central protostar occurs.
In order to speed up the simulations, a density floor is utilized. It
is set up so that the Alfv\'enic time step is not smaller than 
$3\times 10^5\second$. This is a reasonable value in a simulation 
that reaches more than $4\times 10^{12}\second$. 
For similar reasons, the coefficient of ambipolar diffusion is
capped so that its time step is not smaller than $10^6\second$.
In the inner parts of the simulation ($r<1.5\times 10^{15}\cm$)
an additional density floor is set at the relatively low value 
of $2\times 10^{-19}\gm\cm^{-3}$. We monitored the mass added through
the use of the density floors and Alfv\'enic time step limiter and
found it to be insignificant. 

We treat the self-gravity by extending the method of successive
over-relaxation (SOR) used in the two-dimensional (axisymmetric)
core collapse calculations of \citet{Li+2011} to 3D (see also
\ct{Ramsey+2012}). In 2D,
the boundary condition for the gravitational potential was
obtained by direct summation. This proved to be too expensive
in 3D, however. We employed instead the method of multipole
expansion (with degree $l=5$) to determine the boundary condition
on a grid extended by 10 zones in both the inner and
outer radial directions. The new grid extends from 
$r=5\times 10^{13}$ to $2\times 10^{17}\cm$.

\section{Stability of Axisymmetric, Ambipolar Diffusion-Mediated
Protostellar Accretion Flows}
\label{stability}

\begin{figure}[b]
\epsscale{0.6}
\plotone{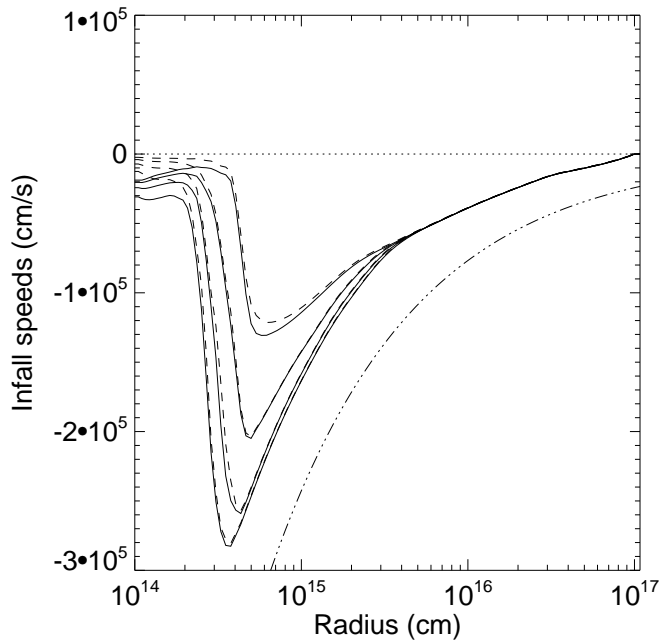}
\caption{Infall speeds in the equatorial region of a representative 2D
(axisymmetric) protostellar accretion flow in the presence of
ambipolar diffusion, along four radial directions that are
(from top to bottom) 0.31, 0.97, 1.68 and 2.44 degrees from the
equatorial plane. The ion and neutral speeds are plotted as dashed
and solid lines, respectively. The free fall speed is plotted for
reference (dashed-dotted).
}
\label{fig:InfallSpeed}
\end{figure}

Ambipolar diffusion is the most widely studied non-ideal MHD effect
in star formation, because it dominates other non-ideal effects at
densities typical of dense cores. It allows the magnetized core
matter to collapse
onto the central object without dragging the field lines along,
as discussed in \S \ref{intro}. Previous 2D (axisymmetric) calculations have
shown that the redistributed magnetic field piles up in a small
circumstellar region, and becomes increasingly dynamically dominant
there (see, e.g., \ct{Li+2011}). This situation is illustrated in
Fig.\ \ref{fig:InfallSpeed}, which shows the infall speeds in
the equatorial region during the collapse of a non-rotating,
magnetized ($\lambda=2.92$) core, in the presence of
ambipolar diffusion ($\zeta=9\times 10^{-17}\second^{-1}$). The
time shown is $t=4.25\times 10^{12}\second$, when $0.12\msun$
has been accreted onto the central object. It
is clear that the equatorial infall is decelerated to a
speed much smaller than the local free-fall speed inside a
radius of roughly $4\times 10^{14}\cm$ (corresponding to the
hydromagnetic shock first studied by \ct{LiMcKee1996}, 
see \S \ref{intro}), where the redistributed
magnetic flux accumulates. In this decelerated region, the material
is held up against the gravity by magnetic forces, a situation
that is prone to instability in 3D.

\begin{figure}[b]
\epsscale{0.6}
\plotone{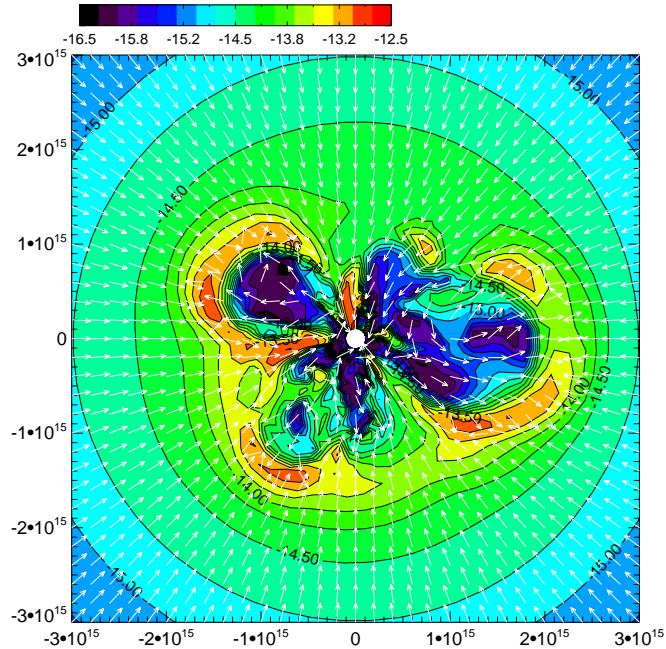}
\caption{Distribution of the logarithm of the mass density $\rho$ (in
${\rm{g}}\cm^{-3}$) and velocity field (unit vectors) on the equatorial
plane, at a representative time for Model A, which is restarted from
the 2D (axisymmetric) calculation at the time shown in
Fig.\ \ref{fig:InfallSpeed}. The length unit is cm.
}
\label{fig:ModelA_color}
\end{figure}

\begin{figure}[b]
\epsscale{0.6}
\plotone{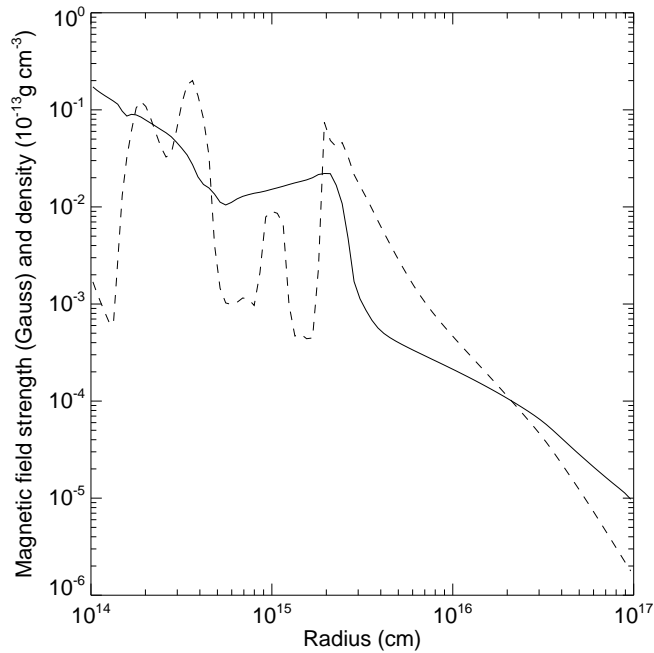}
\caption{Distribution of the total magnetic field strength (solid line)
  and mass density (in units of $10^{-13}\gm\cm^{-3}$, dashed)
along the positive $x$-axis in Fig.\ \ref{fig:ModelA_color}.
}
\label{fig:ModelA_lines}
\end{figure}

To investigate the stability of the magnetically supported structure
induced by ambipolar diffusion in the accretion
flow, we restart, in 3D, the 2D calculation at the time shown in
Fig.\ \ref{fig:InfallSpeed}, when a central object of $0.12\msun$
has already formed (Model A in Table \ref{table:first}). We
find that the axisymmetry is broken quickly, with regions of outward
motion first developing near the inner boundary and then expanding
to larger distances. After a relatively short time of
$2\times 10^{10}\second$, the expansion reaches a size of order
$2\times 10^{15}\cm$ (20 times the radius of the inner boundary),
as shown in Fig.\ \ref{fig:ModelA_color}, where the density distribution
and velocity field
on the equatorial plane are plotted. The figure shows that the
expansion is confined mostly to low-density lobes. These
evacuated lobes are filled with a relatively
strong magnetic field, as illustrated in Fig.\ \ref{fig:ModelA_lines},
where we show the
distribution of the total field strength and the mass density along
the positive $x$-axis in Fig.\ \ref{fig:ModelA_color} (with
$\theta=\pi/2$ and $\phi=0$, cutting through the right lobe).
In the low density region (between
radius $\sim 4\times 10^{14}$ and $\sim 2\times 10^{15}\cm$),
the dynamics is completely dominated by a strong, nearly uniform
magnetic field, with a strength ($\sim 10^{-2}\gauss$) much larger than
at larger radii (outside the magnetically dominated lobe).
The increase in field strength at small radii is associated with the
dense filaments inside the right lobe that are visible in
Fig.\ \ref{fig:ModelA_color}. We have verified that the magnetic
pressures in the evacuated lobes are large enough to overwhelm
the ram pressure of the infalling material. The pressure imbalance
is the reason for the observed expansion in those directions.
The expanding lobes are reminiscent of the so-called ``decoupling
enabled magnetic structure'' (DEMS) found by \citet{Zhao+2011}
in their 3D ideal-MHD AMR simulations of core collapse including
sink particles. As in \citet{Zhao+2011}, the dense structures surrounding
the expanding regions are ring-like rather than shell-like in 3D
(see their Fig.\ 3); they are created out of the dense equatorial
pseudo-disk that is already highly flattened to begin with (see
the top-right panel of Fig.\ \ref{fig:ModelB_color} below).

We have carried out several variants of the above model, including
models with either a ten times weaker initial magnetic field, a nine
times lower rate of cosmic ray ionization, or a non-zero initial
rotation rate. The results are qualitatively similarly, namely,
the initially axisymmetric inner protostellar accretion flow quickly
becomes unstable in the azimuthal direction in 3D. An implication is
that the assumed smooth protostellar accretion flow is unlikely
to be achievable in the first place. We now demonstrate that this is
indeed the case.

\section{Unstable Protostellar Accretion Flows}
\label{filamentary}

\subsection{Reference Model}
\label{reference}

\begin{figure}[b]
\epsscale{1.1}
\plottwo{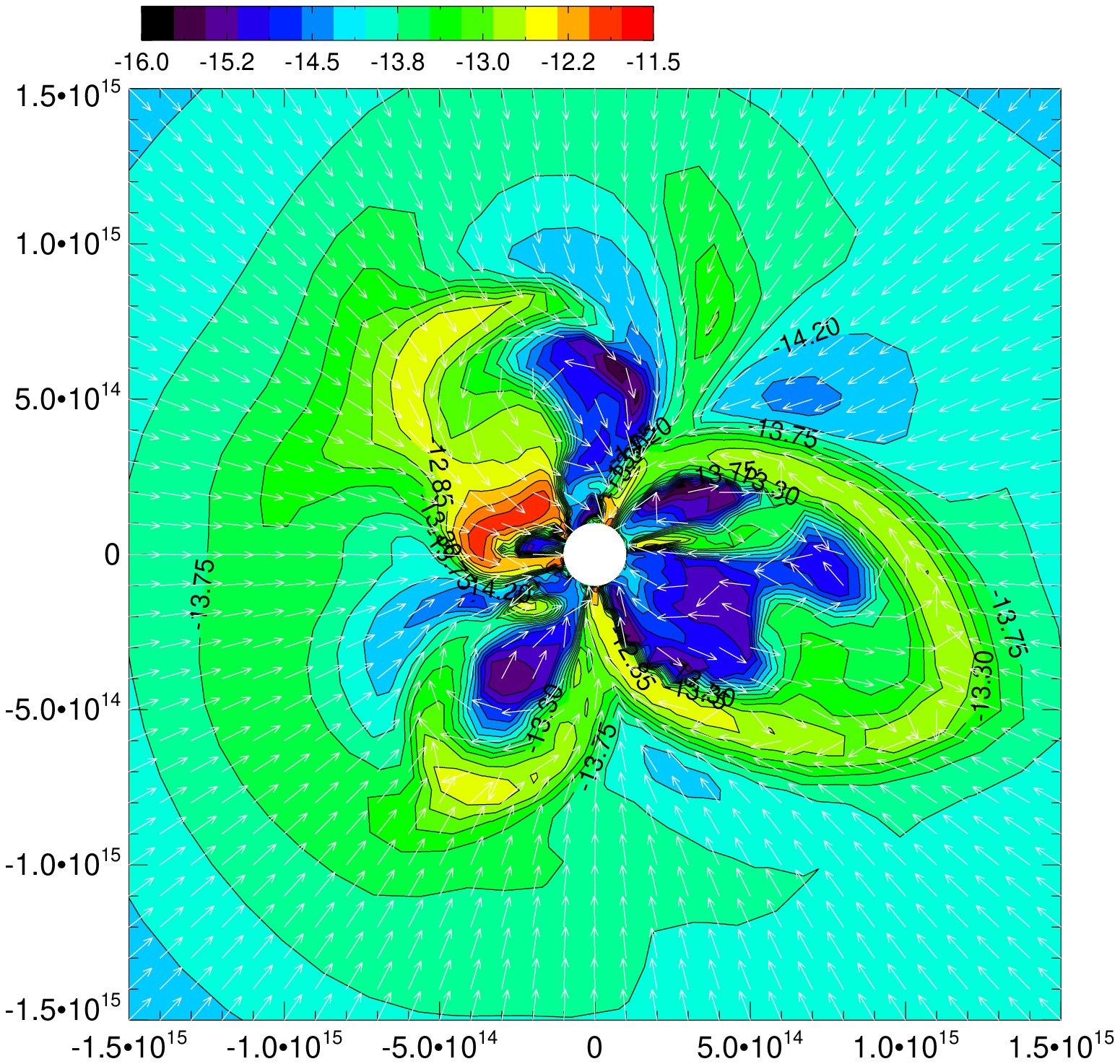}{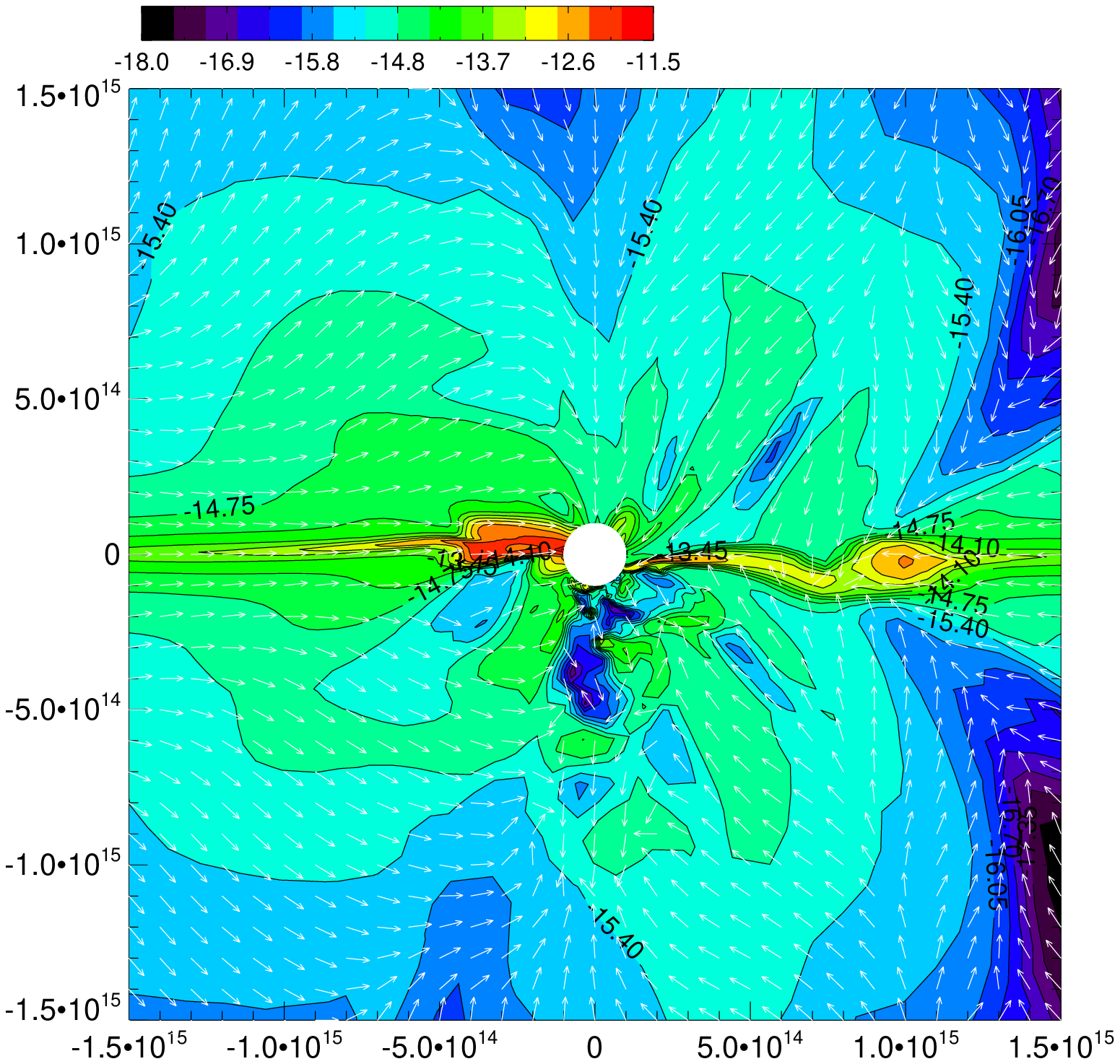}
\plottwo{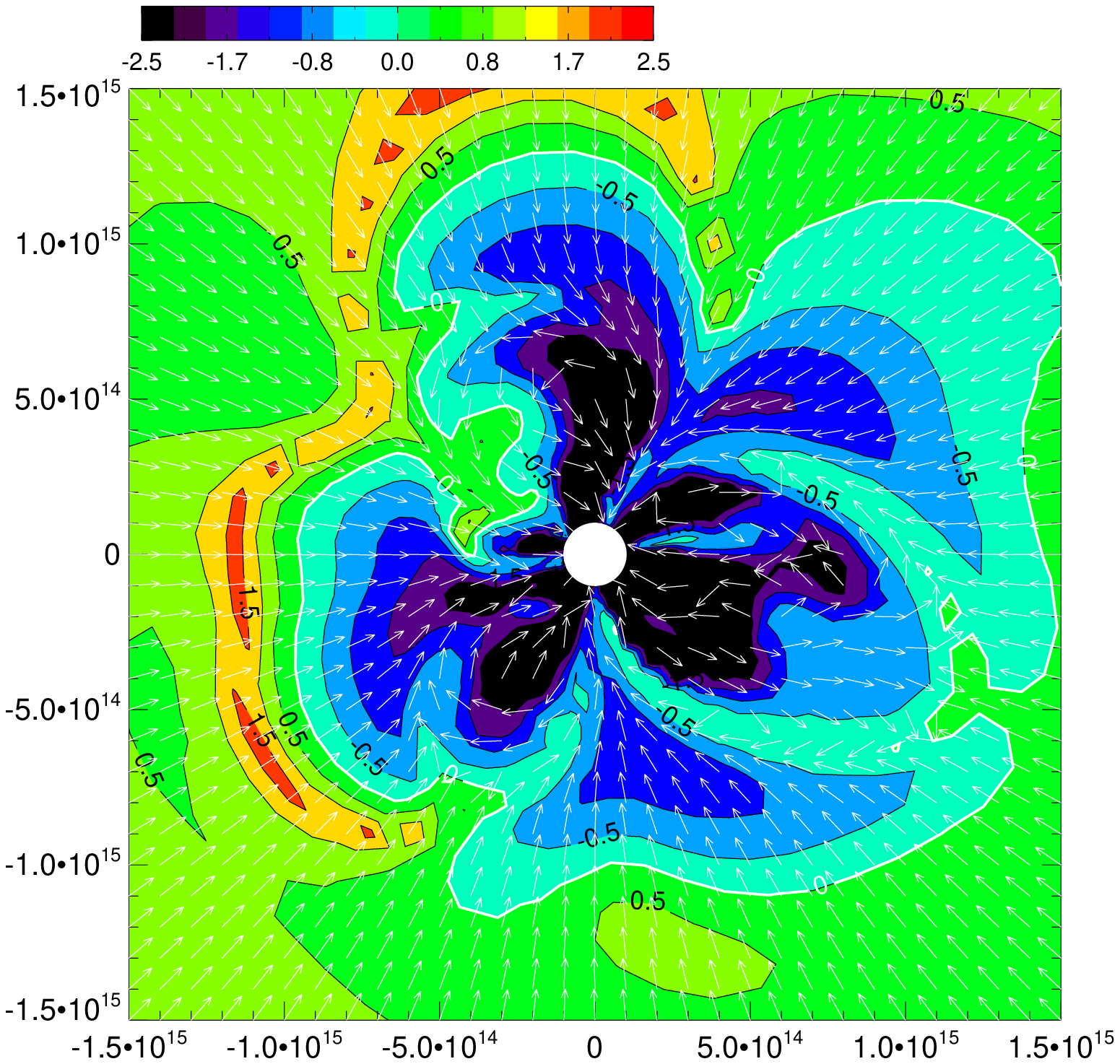}{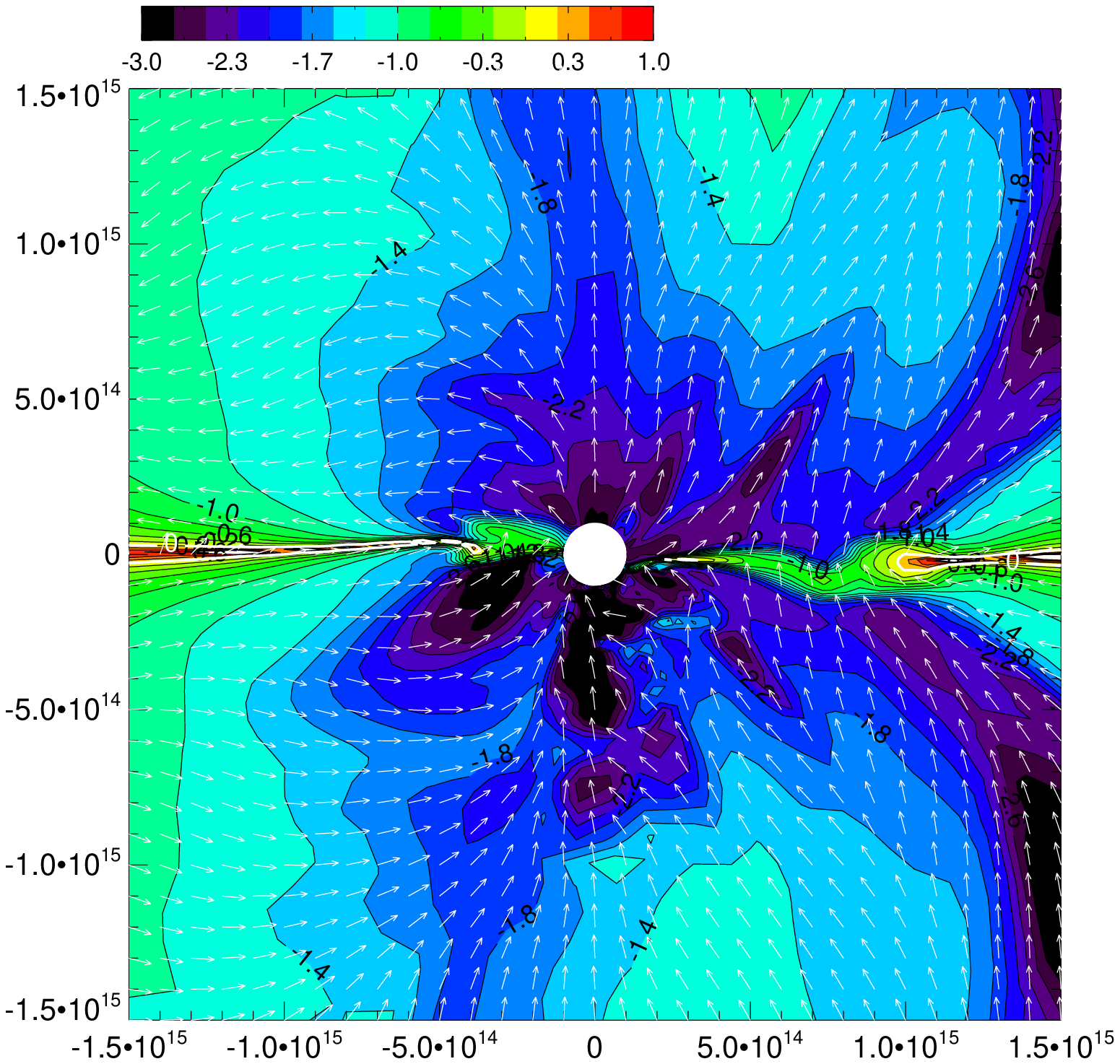}

\caption{Distribution of the logarithm of the mass density $\rho$ and
velocity field (unit vectors) on the equatorial plane (top-left
panel) and a representative meridian plane (top-right) for Model B, at a
time when the central mass is $0.092\msun$. The bottom panels
show the distribution of the logarithm of the plasma-$\beta$ on the
equatorial plane (left, with velocity unit vectors superposed) and
the representative meridian plane (right, with magnetic unit vectors
superposed). The white contours in the two panels mark $\beta=1$.
}
\label{fig:ModelB_color}
\end{figure}

We have restarted the above calculation (Model A) in 3D from the
very beginning ($t=0$), when the core is assumed to be a uniform
sphere. We find that the core remains axisymmetric
during most of the (long) prestellar evolution. To save computation
time, we skip the uneventful early part of the prestellar core
evolution, and restart most of our 3D calculations from 2D
calculations shortly before a central object of significant mass
has formed and any visible asymmetry has developed. In this
subsection, we will concentrate on a representative of such models,
the Model B in Table \ref{table:first}, which serves as a reference for the other
models to compare with. This model is identical to Model A discussed
above, except that we restart the 3D collapse calculation from the
2D collapse at an earlier time of $t=4.12\times 10^{12}\second$,
when the central object contains only a tiny mass of $2.72\times
10^{-5}\msun$ (much smaller than the $0.12\msun$ in
Model A).
We find that the magnetized collapsing flow starts to become visibly
asymmetric during the transition between the prestellar phase of core
evolution to the protostellar phase of mass accretion, when the mass
accretion rate onto the central object increases rapidly. During
the protostellar accretion phase, matter continues to collapse onto
the central
protostar in some azimuthal directions, while the magnetic
flux dragged in by the accretion flow escapes in the other directions,
driving outward motions against the inward collapse. The filamentary
structure resulting
from the tussle between the gravity-driven inflow and the flux
escape-driven expansion is illustrated in the top-left panel of
Fig.\ \ref{fig:ModelB_color}, where we
plot the density distribution and velocity field on the equatorial
plane at a time $t=4.22\times 10^{12}\second$, when
$0.092\msun$  has collapsed into the central object. As emphasized
earlier, the high density regions shown in the panel are not parts
of dense shells. Rather, they are filaments that lie near the
equatorial plane (see the top-right panel of Fig.\ \ref{fig:ModelB_color}),
as part of the magnetically
flattened pseudo-disk that would have formed in 2D but is
disrupted in 3D by the escaping bundles of magnetic field lines.
The flow morphologies displayed in these two panels illustrate the
highly dynamic nature of the inner ($10^2\au$-scale) protostellar
accretion flows that form out of the collapse of dense cores
magnetized to the observed level. The dynamic nature can be seen
more vividly in the movies that can be requested from the authors.

It should not be surprising that the magnetic field is dynamically
important in the inner part of the protostellar accretion flow. In
the bottom two panels of Fig.\ \ref{fig:ModelB_color}, we plot the plasma
$\beta$ (the ratio of thermal
to magnetic pressures) on the equatorial plane and a representative
meridian plane (along $\phi=0$ and $\pi$). It is clear from the
bottom-right panel
that the vast majority of the volume on the $10^2\au$ scale is filled
with low-$\beta$ plasma, as a result of the increase in field strength
due to collapse-induced compression and mass settlement along the
field lines into
the pseudo-disk in the equatorial region. Even on the equatorial
plane, most of the area is covered by strongly magnetized material
with $\beta \ll 1$ within a radius of $10^2\au$ at the time shown,
although there are ``fingers'' of less strongly magnetized material
with $\beta \sim 1$ (see the bottom-left panel).
It is these dense, less magnetized ``fingers'' or
``filaments'' that dominate the mass accretion onto the central
object, which is at a rate of $\sim 3\times
10^{-5}\msun\yr^{-1}$ at the time shown. The infall material
percolates through a ``sea'' of highly magnetized, low-density medium.

\begin{figure}[b]
\epsscale{1.1}
\plottwo{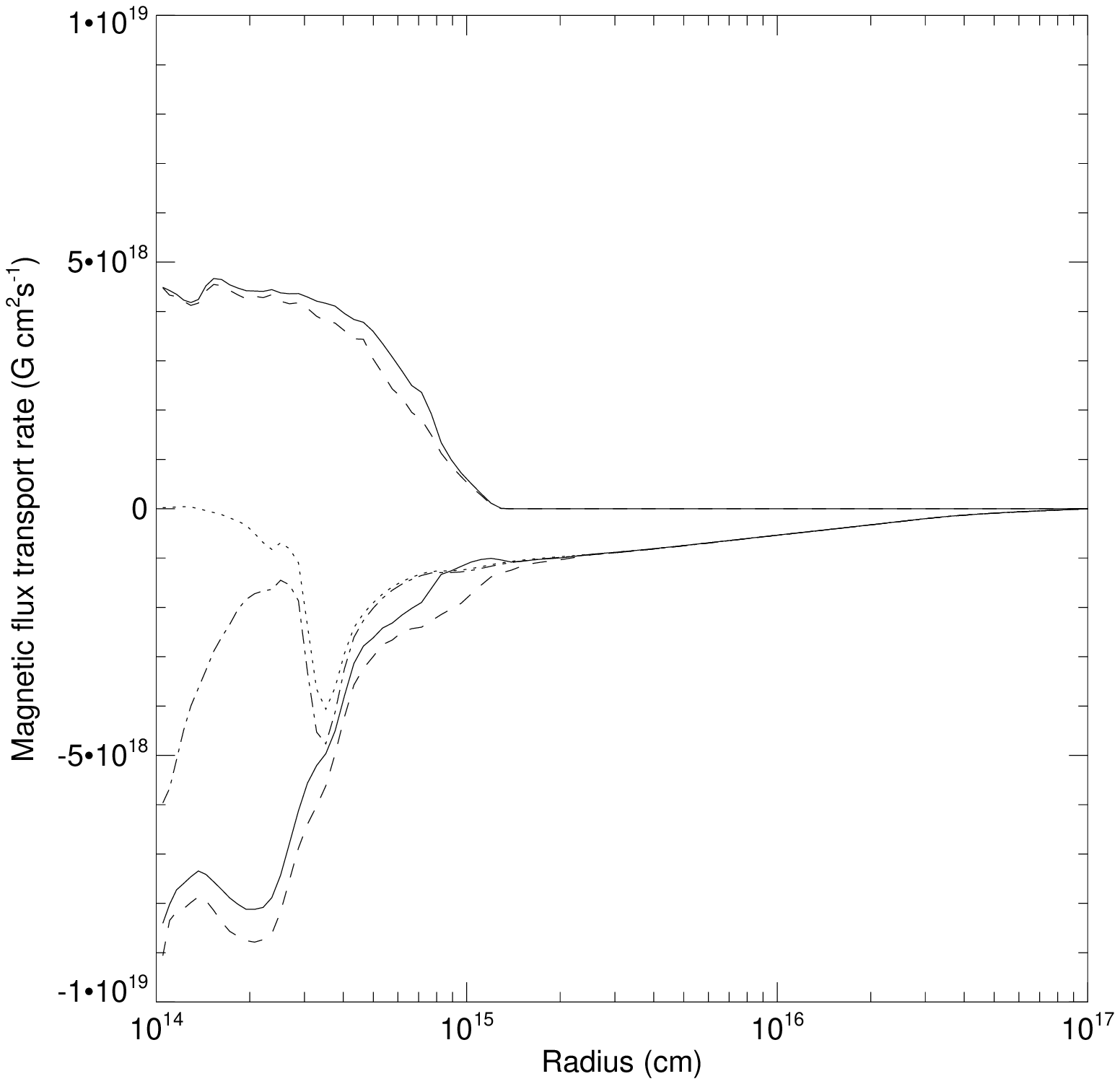}{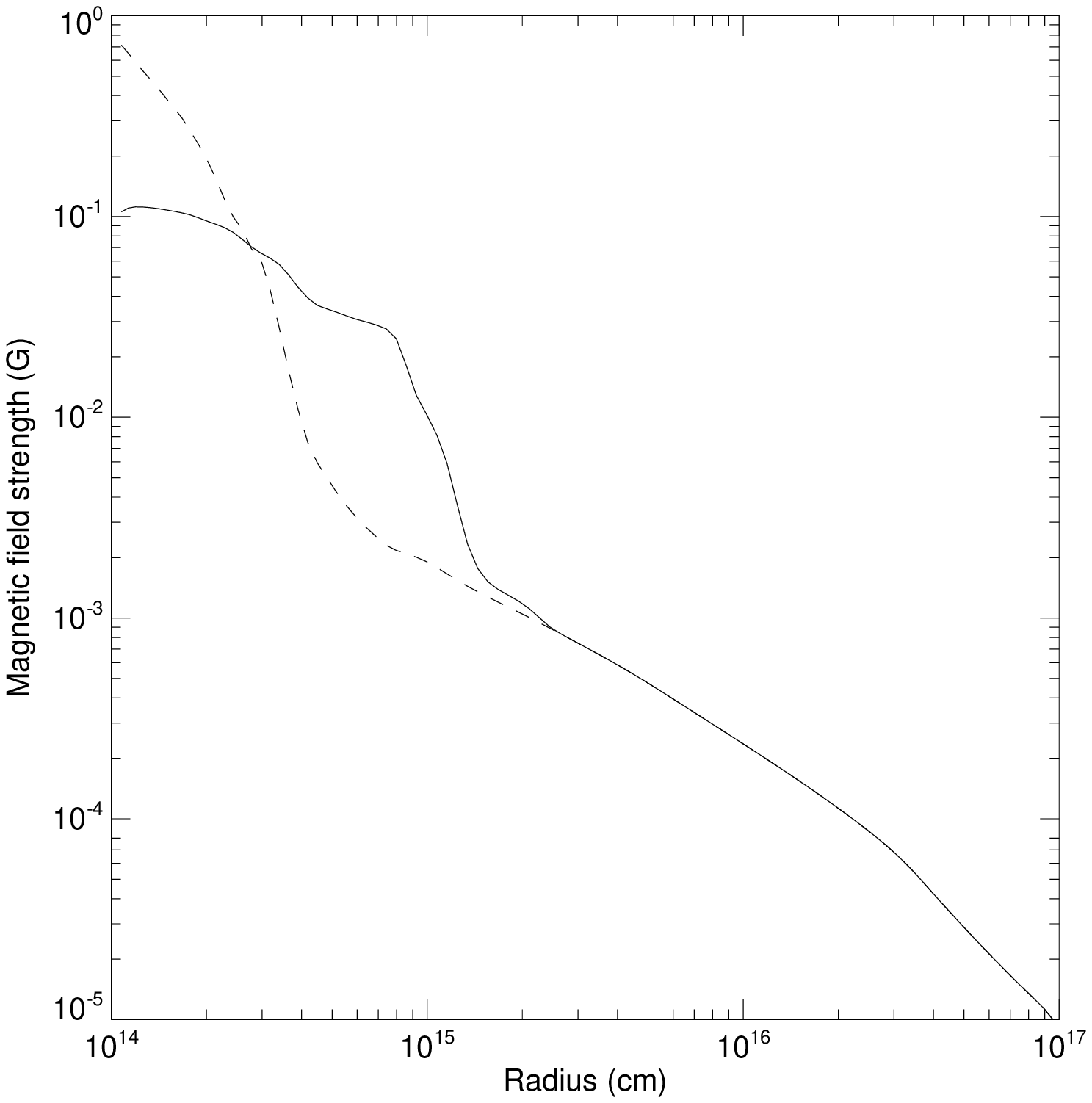}
\plottwo{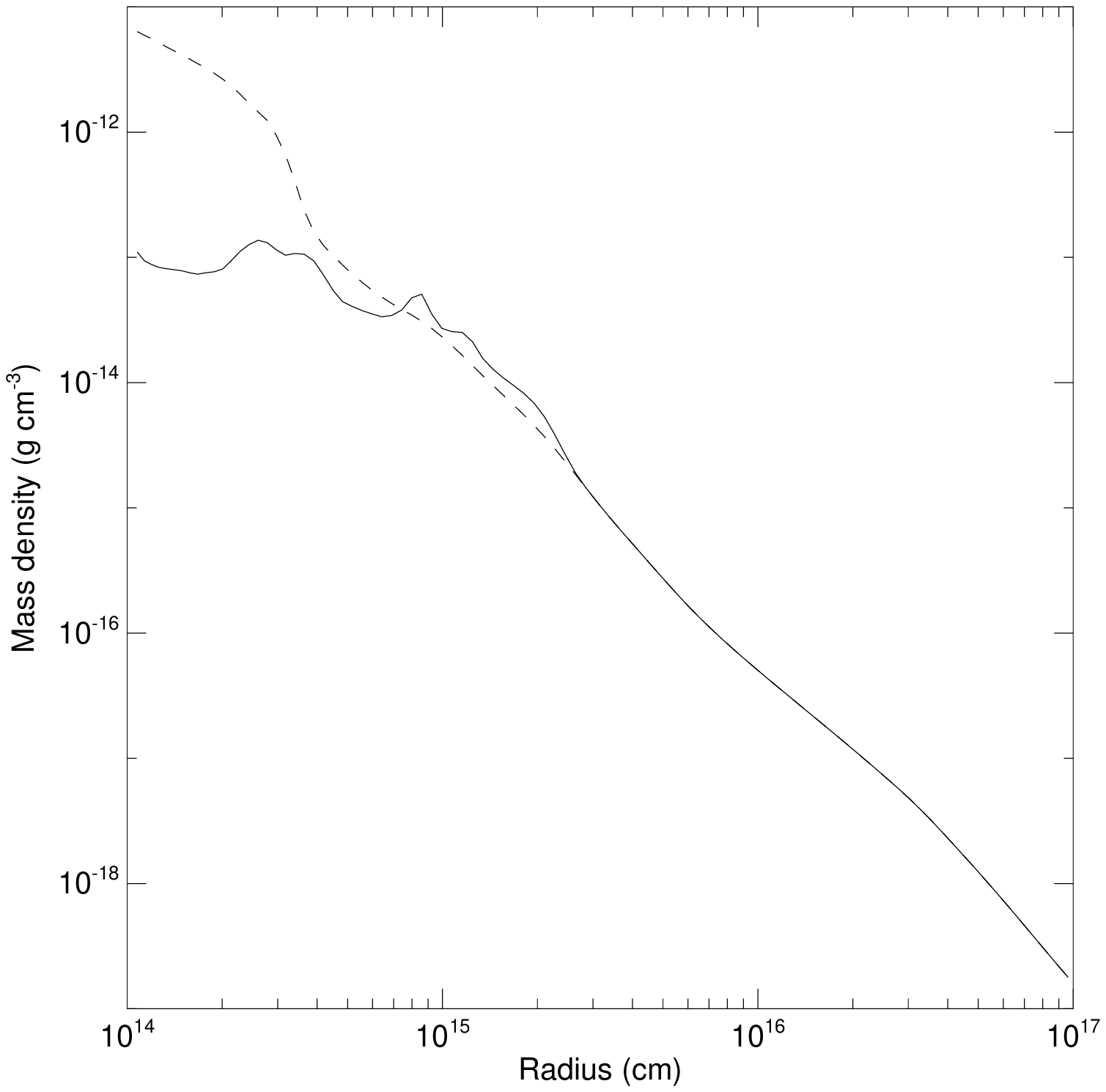}{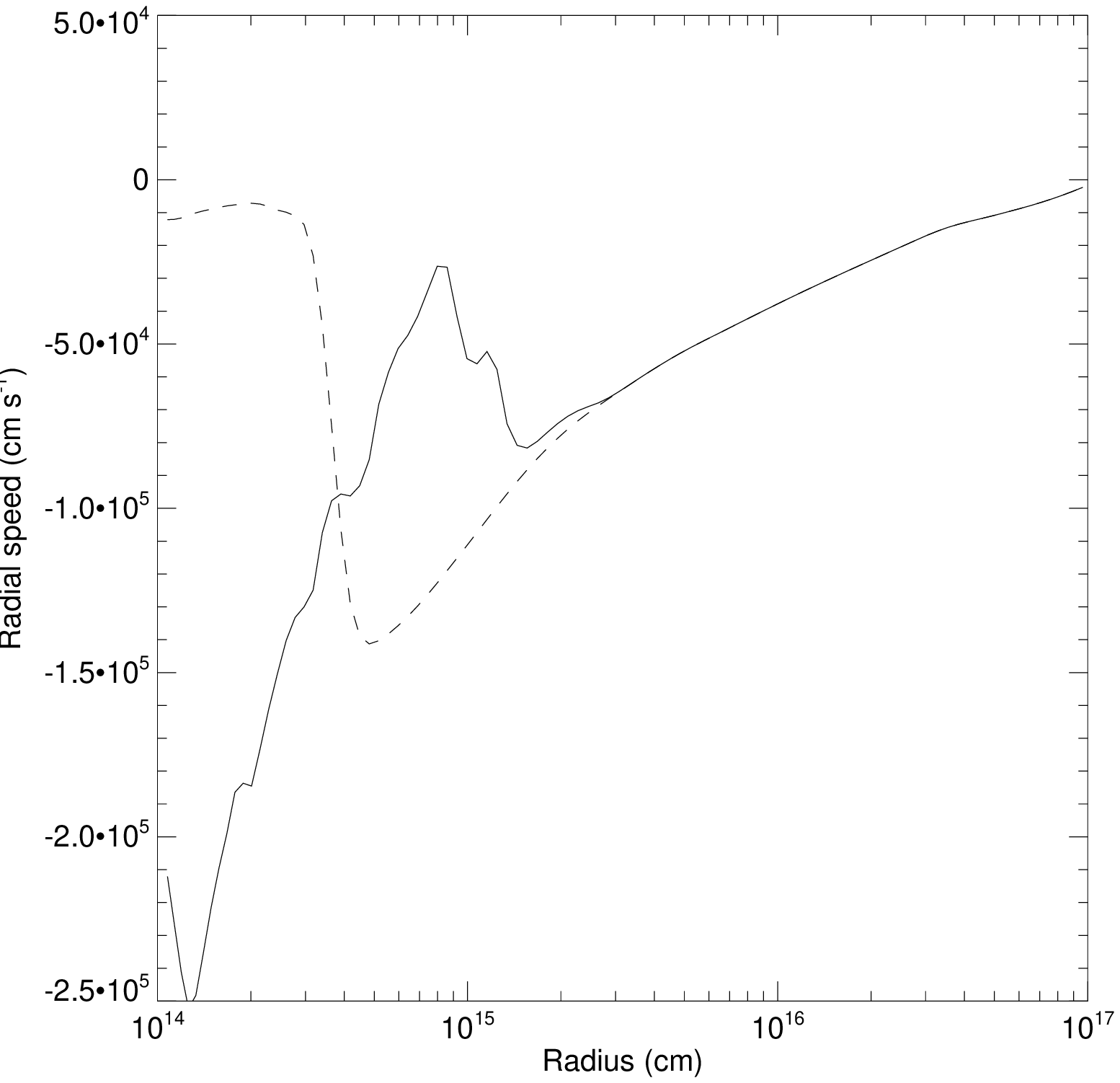}

\caption{Top-left panel: rates of magnetic flux transport (see 
 equation \ref{fluxrate})
  associated with the
  ion (${\dot \Phi}_i$, solid lines) and neutral (${\dot \Phi}_n$,
dashed) velocities across circles of
different radii on the equator for (3D) Model B. The bottom and top
  pairs of lines are for the inward and outward flux
transport, respectively. The rates of flux transport associated with
  the ion (dotted line) and neutral (dot-dashed) velocities in 2D are also
  shown for comparison. The remaining panels compare the azimuthally
  averaged radial distributions of the vertical field strength 
(top-right panel), mass
  density (bottom-left), infall speed weighted by mass (bottom-right)
  on the equatorial plane for the 3D (solid lines) and 2D (dashed)
models.
}
\label{fig:ModelB_lines}
\end{figure}

It is instructive to compare the 3D simulation shown in Fig.\ \ref{fig:ModelB_color}
more quantitatively to the 2D version of the simulation at the same
time ($t=4.22\times 10^{12}\second$). One fundamental difference between
the 3D and 2D models is how the magnetic flux is transported. In the
top-left panel
of Fig.\ \ref{fig:ModelB_lines}, we display the rate of magnetic flux
transport, $\dot \Phi$, across a circle $C$ of radius $r$ and
circumference $\partial C$ on the
equatorial plane, computed as
\begin{equation}
{\dot \Phi}_i \equiv
\frac{\partial \Phi}{\partial t}
=-\int_C\frac{\partial {\bf B}}{\partial t}\cdot d{\bf S}
=\int_C\nabla\times({\bf B}\times{\bf v}_i)\cdot d{\bf S}
=\int_{\partial C}({\bf B}\times{\bf v}_i)\cdot d{\ell}
=\int_0^{2\pi}(B_r v_{\theta,i}-B_\theta v_{r,i})\,r\,d\phi
\label{fluxrate}
\end{equation}
where the subscript ``i'' stands for ions, which are assumed to be
well coupled to the magnetic field, $d{\bf S}$ and $d{\ell}$
are area and length vector elements, and $\Phi$ is the
flux.\footnote{The four equal signs in equation (\ref{fluxrate}) are justified
based on the definition of magnetic flux, the induction equation,
Stokes' theorem, and expansion on spherical coordinates,
respectively. The definition of magnetic flux is based on $B_z=-B_\theta$.  }
For comparison, we also compute the
rate of flux transport associated with the neutral velocity
${\dot \Phi}_n$; it is the
expected rate in the absence of ambipolar diffusion.
We find that, in 2D, the bulk neutral material would have dragged in
magnetic flux at a high rate (the dot-dashed line in the panel) at small
radii were it not for ambipolar diffusion. The ion-neutral drift
reduces the rate greatly, especially
close to the inner boundary (see the dotted line).
In contrast, the ion-neutral drift is much less
effective in reducing the rate of inward magnetic flux transport
in 3D (mostly by infalling material), as seen from the small separation
between the bottom solid and dashed lines. Most of the magnetic
flux carried in by the infalling neutral material is advected back
out by the low-density expanding material; the outward flux advection
is not modified by ambipolar diffusion much either (note the small
separation between the top solid and dashed lines). The conclusion is
that, in 3D, the magnetic transport is dominated by advection through
bulk fluid motions, with ambipolar diffusion playing a much reduced
role compared to the 2D case. We will return to this important
conceptual point in the discussion section (\S \ref{modes}).

The different modes of magnetic flux transport in 2D and 3D have
profound effects on the dynamics of the protostellar accretion
flow. In the remaining three panels of Fig.\ \ref{fig:ModelB_lines}, we compare
the azimuthally averaged radial distributions of the vertical field
strength, mass density, and infall speed on the equatorial plane for
the 2D and 3D models. It is clear that the magnetic field is more
concentrated at small radii in 2D than in 3D (see the top-right
panel), because the microscopic
ambipolar diffusion in 2D is less efficient in smoothing out the field
concentration than the macroscopic flux advection in 3D. The reduction
in the field concentration is the main reason why ambipolar diffusion
plays a reduced role in the magnetic flux transport in 3D, as pointed
out above. The reduction in magnetic field strength and the associated
magnetic forces in 3D enables the accretion flow to collapse faster
toward the central object (see the bottom-right panel), which in turn
leads to lower densities at small
radii (see the bottom-left panel). In 3D, a larger region is affected
by the accreted magnetic flux, which is more widely redistributed.

\subsection{Rotation, Ionization Level, and Ohmic Dissipation}

\begin{figure}[b]
\epsscale{1.1}
\plottwo{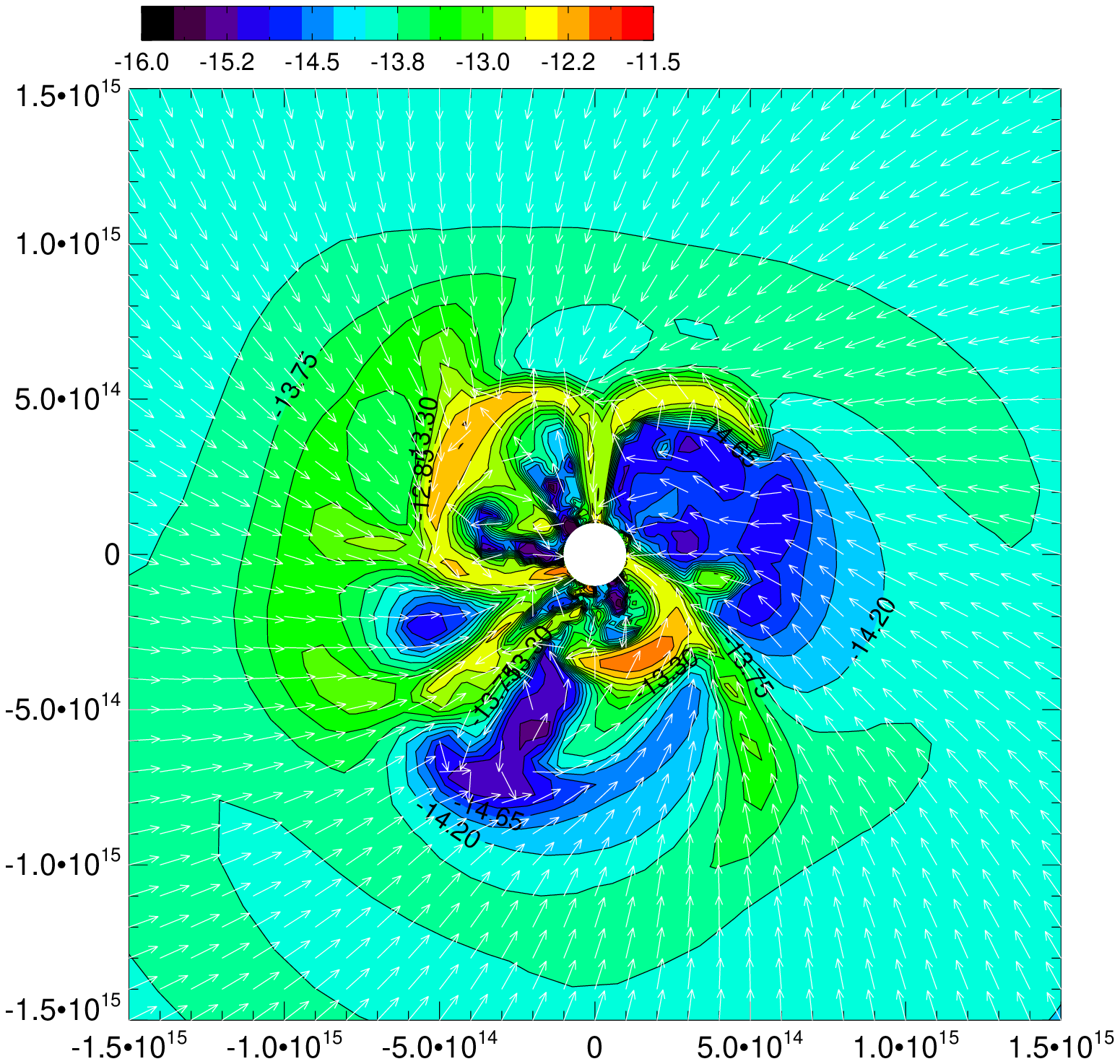}{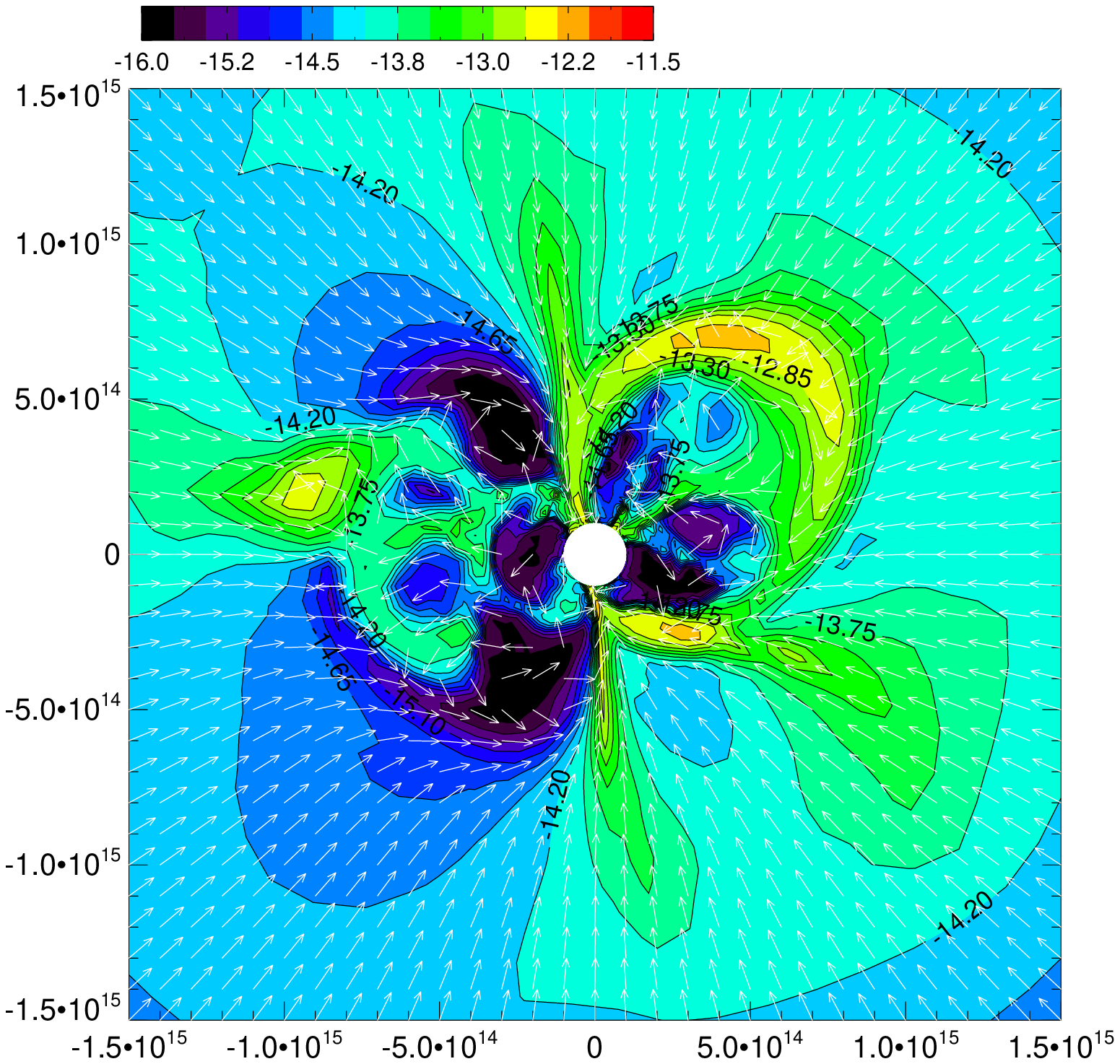}
\plottwo{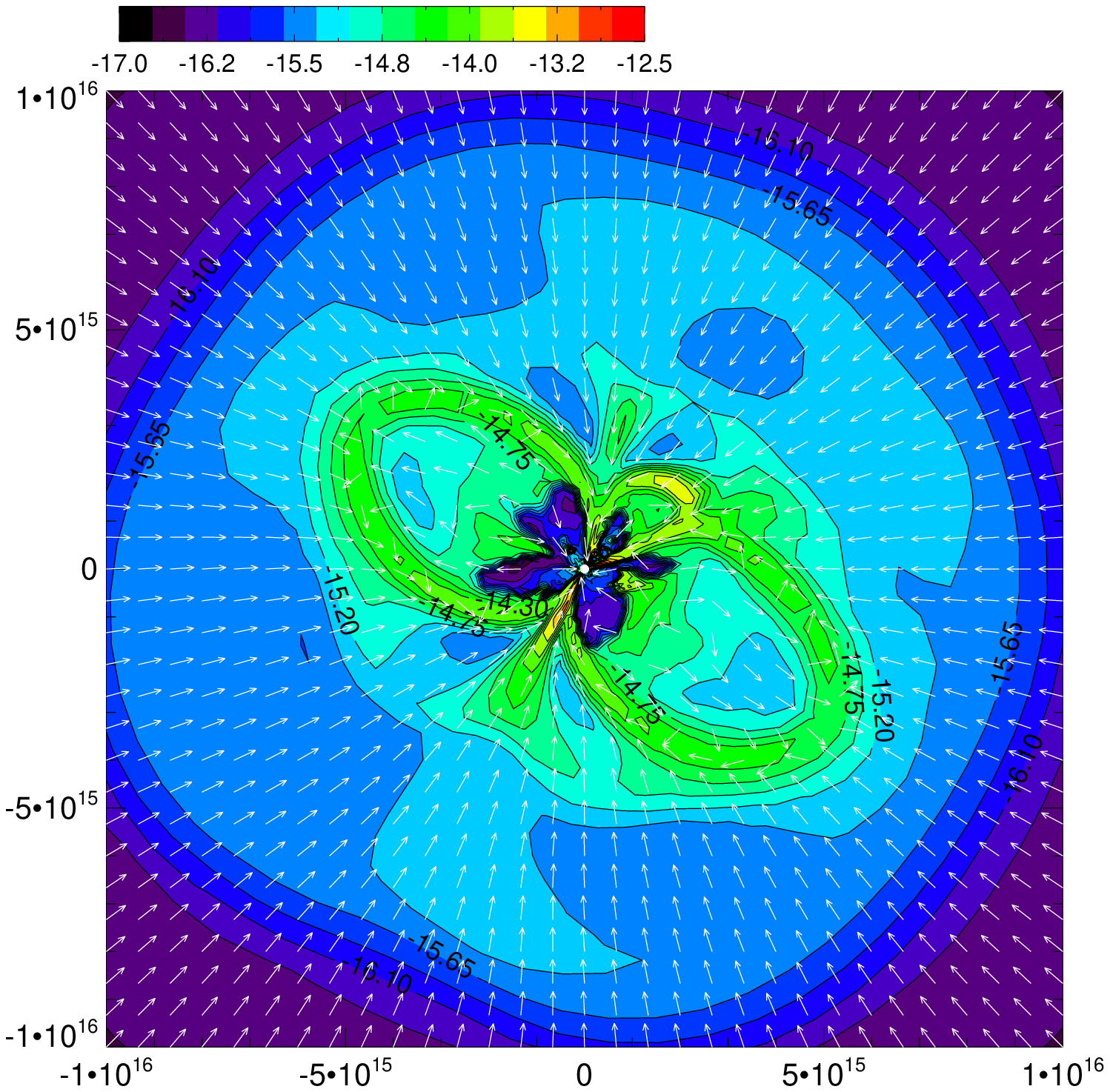}{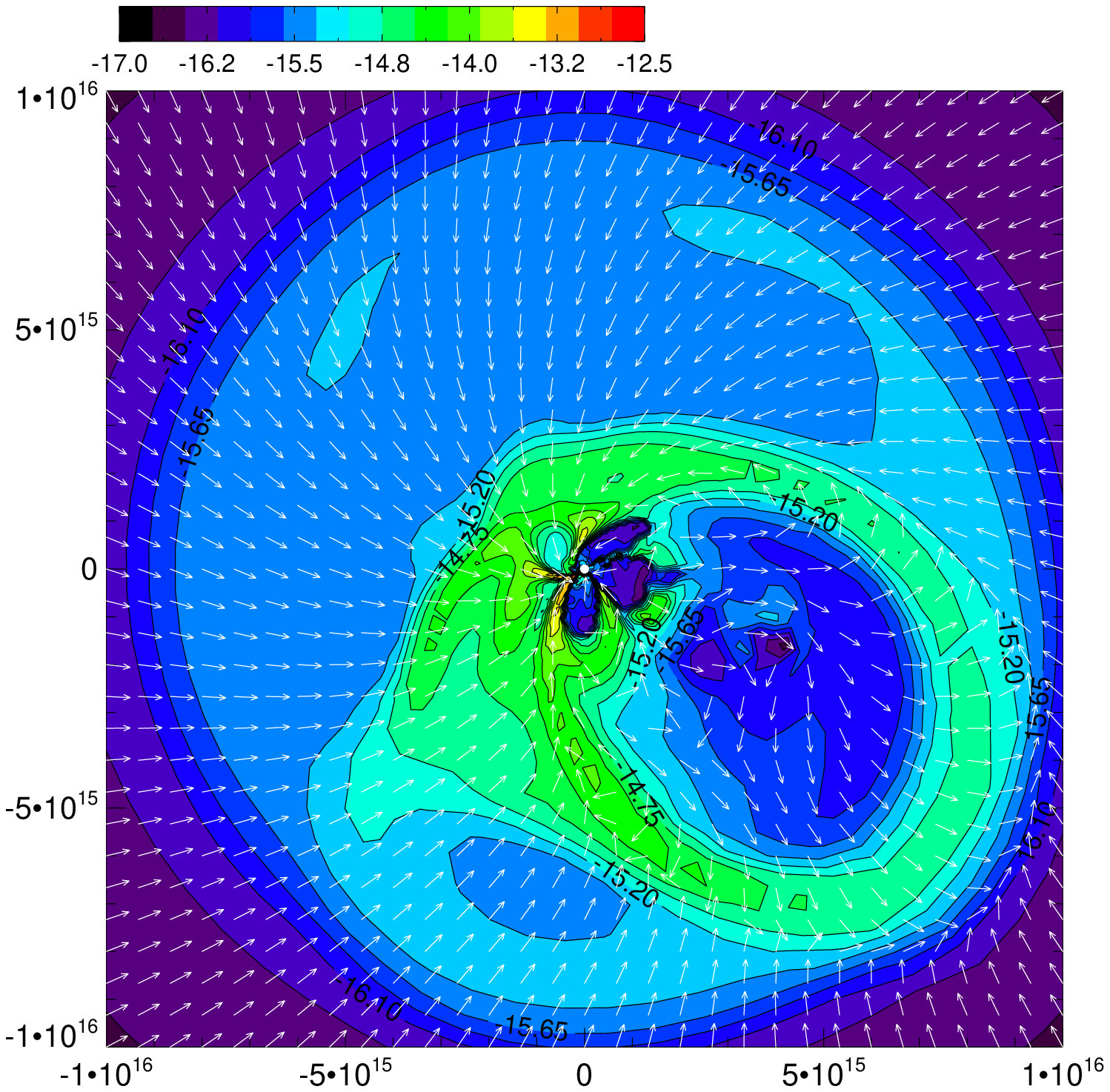}
\caption{Distribution of the logarithm of the mass density $\rho$ and
velocity field (unit vectors) on the equatorial plane, for Models
C (top-left panel), D (top-right), E (bottom-left) and F (bottom
right), all showing filamentary structures in the inner protostellar
accretion flow. Note the change in size scale between the top and bottom
panels.
}
\label{fig:Models_CDEF}
\end{figure}

Unstable, filamentary protostellar accretion is not unique to the
reference model (Model B) that includes ambipolar diffusion. We
have carried out dozens of runs with different model parameters
and different non-ideal MHD effect (Ohmic dissipation), and they
all show a qualitatively similar behavior. Fig.\ \ref{fig:Models_CDEF} displays
four examples. In the top-left panel, we plot the velocity field and
density distribution on the equatorial plane of a
case that is identical to the reference run except for the initial
rotation rate, which is now $\Omega=10^{-13}\second^{-1}$ rather than
zero (Model C in Table \ref{table:first}). The counter-clockwise rotation can be
seen in the panel. It does not fundamentally change the
filamentary morphology of the accretion flow. In particular, a
rotationally supported disk (RSD) has not formed up to the time
shown ($t=4.575\times 10^{12}\second$), which corresponds to a
relatively early phase of protostellar mass accretion, when the
central mass is only $0.071\msun$. We were unable to run the
simulation much longer because of numerical difficulties associated
with strong magnetic fields in low density regions. The top-right
panel displays Model D, which is identical to the reference Model B,
except for the cosmic ray ionization rate, which now has the
canonical value $\zeta = 10^{-17}\second^{-1}$ instead of $9\times
10^{-17}\second^{-1}$. The snapshot is taken at time $t=4.163\times
10^{12}\second$, when the central mass is $0.096\msun$. Again,
we find filamentary structures shaped by the interplay between
gravitational infall and the magnetically driven expansion.

The unstable, filamentary accretion is not limited to magnetized
collapse in the presence of just ambipolar diffusion. In the bottom
panels of Fig.\ \ref{fig:Models_CDEF}, we show two representative models
with a spatially constant resistivity of
$\eta=10^{17}\cm^2\second^{-1}$; this value is larger than the classical
microscopic value for the density range under consideration, and
is adopted for illustrative purposes only. We have experimented with
$\eta=10^{16}$ and $10^{18}\cm^2\second^{-1}$ and found qualitatively
similar results. The adopted resistivity has the advantage of
enabling the simulations
to run longer in the protostellar mass accretion phase compared
to the ambipolar diffusion cases. The bottom-left panel shows a
non-rotating case that is the same as the reference model, except
that ambipolar diffusion
is now replaced by Ohmic dissipation (Model E), at a time $t=4.5\times
10^{12}\second$, when $0.22\msun$ of mass has been accreted onto
the central object. By this time, the
filamentary accretion region has expanded beyond $\sim 4\times
10^2\au$. The last panel displays a case that is the same as
Model E, except that the core rotates initially with an angular
speed of $\Omega=10^{-13}\second^{-1}$ (Model F in Table \ref{table:first}). As in
the ambipolar diffusion case, there is no hint of the formation of
a rotationally supported disk out of the collapse of the rotating
core, even at the rather late time shown ($t=4.82\times 10^{12}\second$),
when the central mass has grown to $0.16\msun$. We conclude that
protostellar accretion flows are unstable and become filamentary
in the presence of a moderate level of Ohmic dissipation, with or
without rotation, as is true for the models with ambipolar
diffusion.

\subsection{Magnetic Decoupling and Nature of the Instability}
\label{step}

The magnetic field is expected to decouple from the bulk neutral
material sooner or later as the density increases, because the gas
becomes less ionized and the charged particles less well tied to
the field lines. The exact value for the decoupling density is
somewhat uncertain. \citet{Nakano+2002} estimated a value
of a few times $10^{11}\cm^{-3}$. It may however be an order
of magnitude higher according to \citet{KunzMouschovias2010}. Treating the
decoupling fully would require a detailed calculation of the number
densities of all charged species (including dust grains), which is
beyond the scope of this work. Nevertheless, as long as the decoupling
occurs at a high enough density (or close enough to the origin,
as found by \ct{Nakano+2002} and \ct{KunzMouschovias2010}), the
essence of the process is already captured in our simulations, through
the use of an inner radial boundary at $r=10^{14}\cm$ (or $6.7\au$):
matter that crosses the boundary is accreted onto the central object,
and becomes decoupled from the magnetic field lines that were
originally attached to the matter but are now left behind in the
computational domain. The basic features of the
protostellar accretion flow do not depend on the size of the inner
boundary. For example, we have shrunk that size by a factor of 2 for
Model B and E, and found that the flow pattern remains qualitatively
similar.

The models discussed so far rely on the use of an inner boundary
for treating the magnetic decoupling. In Models G and H of
Table \ref{table:first}, we refine the treatment by including a small diffusive
region outside the inner boundary. It is done through a step
function for the resistivity, with $\eta=10^{19}\cm^2\second^{-1}$
inside $r_c=2\times 10^{14}\cm$ (twice the radius of the inner
boundary), and $\eta=1\cm^2\second^{-1}$ outside. The outside
resistivity is so small that the field lines are essentially
frozen in the matter. Inside $r_c$, one would ideally like to
choose an $\eta$ as large as possible, so that the magnetic
field is completely decoupled from the
matter and can thus be easily redistributed relative to the
matter. However, the larger the resistivity $\eta$ is,
the smaller the time step $dt$ must be in order to ensure numerical
stability for our explicit treatment of the Ohmic dissipation.
As a compromise, we settled on a value $\eta=10^{19}\cm^2\second^{-1}$,
which is large enough to illustrate the effects of magnetic
decoupling but small enough that the simulation can be completed
in a reasonable amount of time. We also increased the critical
density for stiffening the equation of state by a factor of
$10^3$, to $\rho_c=10^{-10}\gm\cm^{-3}$, so that the accretion
flow remains isothermal, which makes it easier to see how the
magnetic flux redistribution-driven instability develops.

\begin{figure}[b]
\epsscale{1.10}
\plottwo{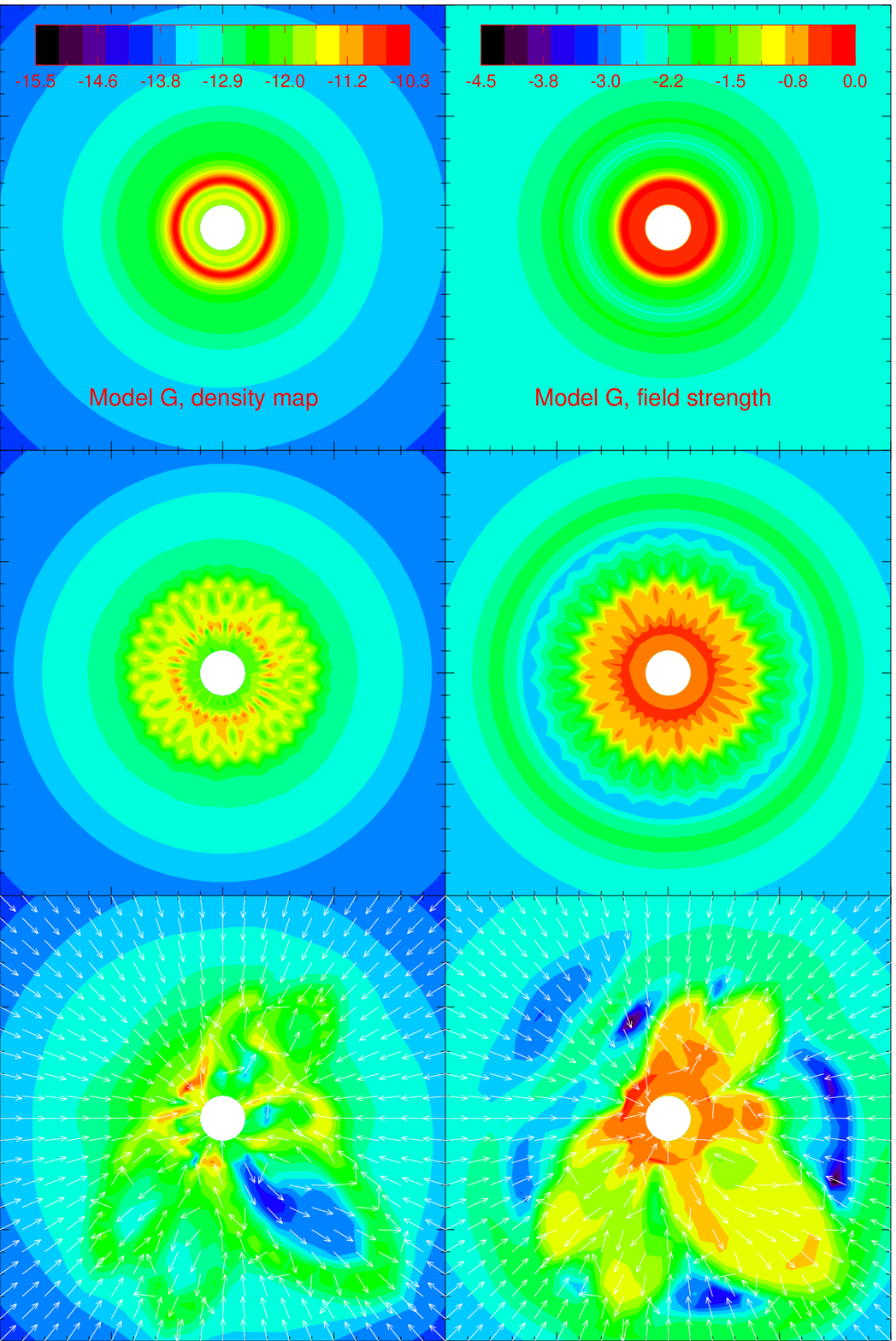} {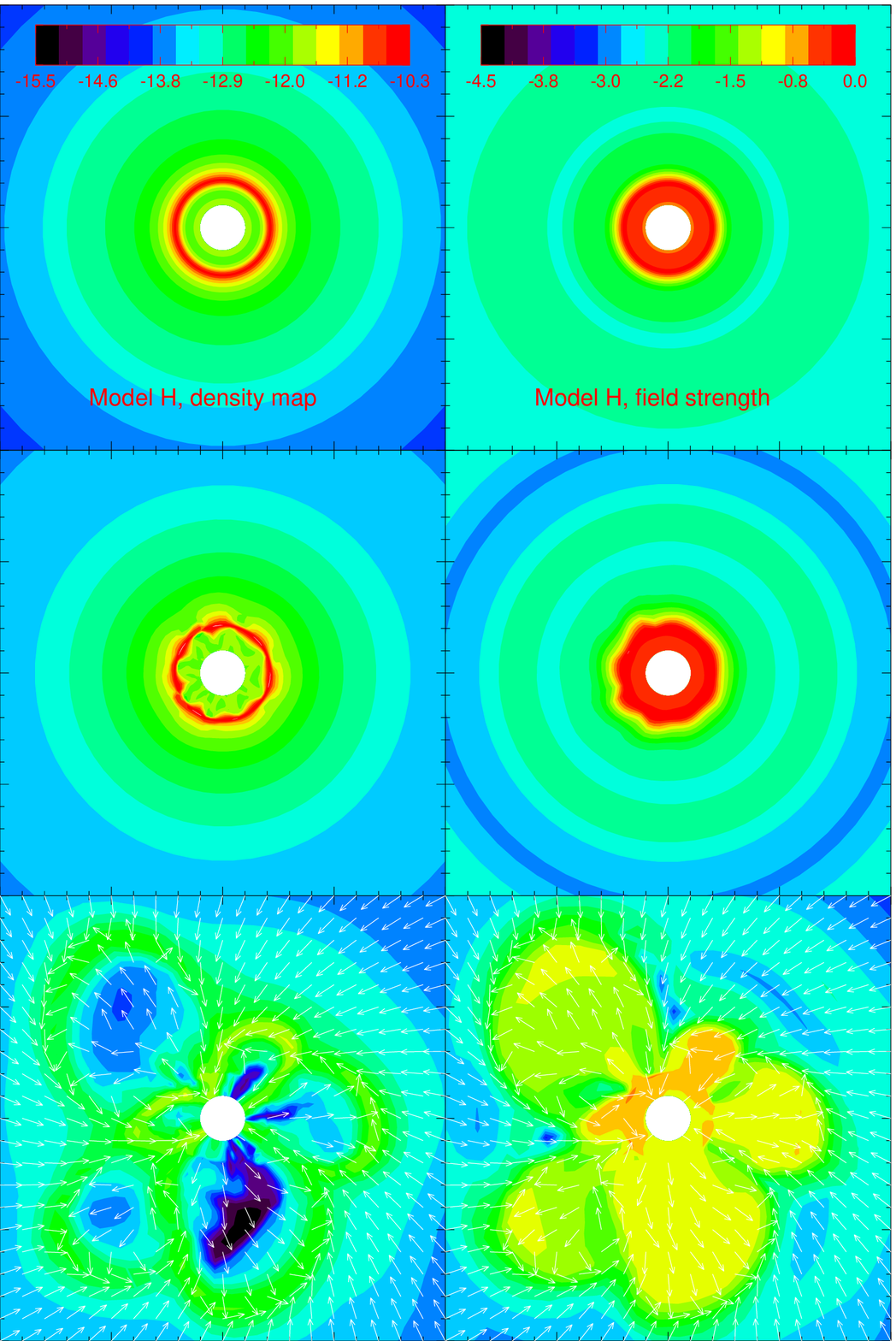}

\caption{Evolution of Model G (non-rotating, left two columns) and H
  (rotating, right two columns), at three representative times
  (from top to bottom). The first and third
  columns show the equatorial density distribution for Model G and H
  respectively, and the second and fourth columns show the
  distribution of the component of magnetic field perpendicular to
  the equator ($B_z=-B_\theta$) for Model G and H. In both models, the
  redistribution of magnetic flux inside the resistive region gives
  rise to interchange instability that leads to advective magnetic
  flux transport, as indicated by the (unit) velocity vectors shown in
  the bottom panels. The size of each panel is $2\times 10^{15}\cm$.}
\label{fig:Models_GH}
\end{figure}

The evolution of Models G and H is illustrated in Fig.\ \ref{fig:Models_GH},
which shows snapshots of the density and field strength distributions
on the equatorial plane at three representative times. We will
concentrate on the non-rotating Model G first. At the earliest time
shown ($t=4.205\times 10^{12}\second$), the accretion flow
remains nearly axisymmetric. The most striking feature is that the
strength of the magnetic field inside the resistive region (within
radius $r_c=2\times 10^{14}\cm$) is more than an order of magnitude
higher than that outside (top row, second column). This is because
magnetic flux is dragged into the resistive region by accretion, and
this magnetic flux is not destroyed by the large local resistivity,
despite occasional claims to the contrary in the literature;
rather, it accumulates in the resistive region. At the time shown
in the top panels, the accumulated magnetic flux is
$\Phi_c=4.95\times 10^{28}\gauss\cm^2$. The total mass inside the
region is $M_c=9.16\times 10^{31}\gm$, with most of the
contribution coming from the central mass, which has
$8.82\times 10^{31}\gm$. The dimensionless mass-to-flux ratio for
the region is therefore $\lambda_c=3.00$, which is close to the
average value for the dense core as a whole ($2.92$), indicating
that the magnetic flux associated with the mass that has entered
the central object is indeed trapped in the resistive region. The value
$\lambda_c=3.00$ is somewhat smaller than the initial value on the
central flux tube (4.38), as expected, because not all of the matter
along the redistributed field lines has collapsed into the resistive
region.

The rapid increase in field strength across the boundary $r_c$ between
the nearly ideal MHD and resistive region implies a large magnetic
pressure gradient near $r_c$, which opposes the local
gravitational collapse. The magnetic pressure force is aided by the
magnetic tension force near the boundary, where the poloidal field
lines become highly pinched. The net effect is a rapid deceleration of the
collapsing flow near the boundary, which leads to a local pile-up of
material. The pile-up corresponds to the density peak near
$r_c$ (see the top-left panel). Inside $r_c$, the magnetic field is less
well coupled to the matter, which allows the gravity to re-accelerate
the gas to high speed and thus lower the density.

The magnetically supported region becomes unstable in the azimuthal
direction shortly afterward, with high frequency modes dominating
initially (see the middle row, left two panels). This is
characteristic of magnetic interchange instability, and has been
seen, for example, in the simulations of accretion disks threaded
by a strong magnetic field by \citet{StehleSpruit2001}. In the
absence of rotation, \citet{SpruitTaam1990} find that the
criterion for the instability is that the mass-to-flux
ratio decreases in the direction of the gravity. This condition is
satisfied in our case because of the magnetic flux redistribution
inside the resistive region, which reduces the mass-to-flux ratio
inside the region compared to that outside.

As the interchange instability grows, lower frequency azimuthal modes
are expected to become more prominent (see, e.g., Fig.\ 3 of
\ct{StehleSpruit2001}). This is indeed the case for our model, as
shown in the bottom row (left two columns) of Fig.\ \ref{fig:Models_GH},
where the distributions of the density and magnetic field strength
are dominated by several lobes, especially the one along the lower-right
direction. This lobe grows preferentially relative to the others. It
dominates the dynamics of the inner accretion flow at later times.

The evolution of Model H with rotation (the right two columns of
Fig.\ \ref{fig:Models_GH}) is similar to the non-rotating Model G.
Magnetic flux is again trapped in the resistive region, which
leads to interchange instability. The instability develops
more gently compared to the non-rotating case, presumably
because it is weakened somewhat by differential rotation, as
predicted from the linear analysis (\ct{LubowSpruit1995}).
The rotation does not fundamentally change the nonlinear outcome of
the instability, however. In both cases, lobes of highly
magnetized material expand away from the origin, transporting
magnetic flux to large distances well outside the small
resistive region. We conclude that the field-matter decoupling
in the resistive region has driven the inner protostellar accretion
flow unstable, which leads to the new mode of advective magnetic
flux transport that does not depend on local microscopic magnetic
diffusion and that operates even in the ideal MHD part of the
flow. This is in agreement with the ideal MHD simulations of
\citet{Zhao+2011}, where the magnetic decoupling is represented with a
sink particle treatment.

\begin{figure}[b]
\epsscale{1.1}
\plottwo{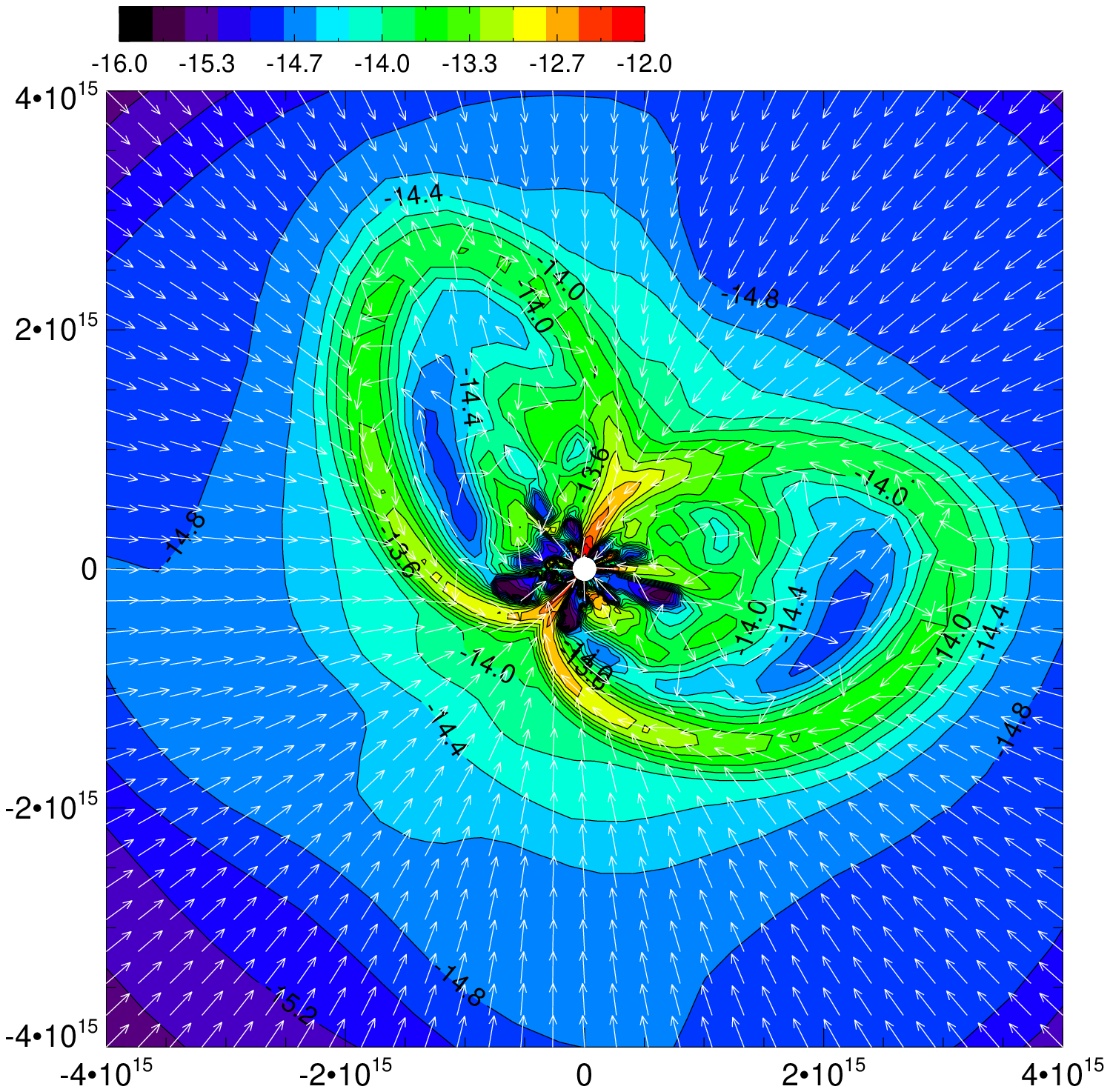}{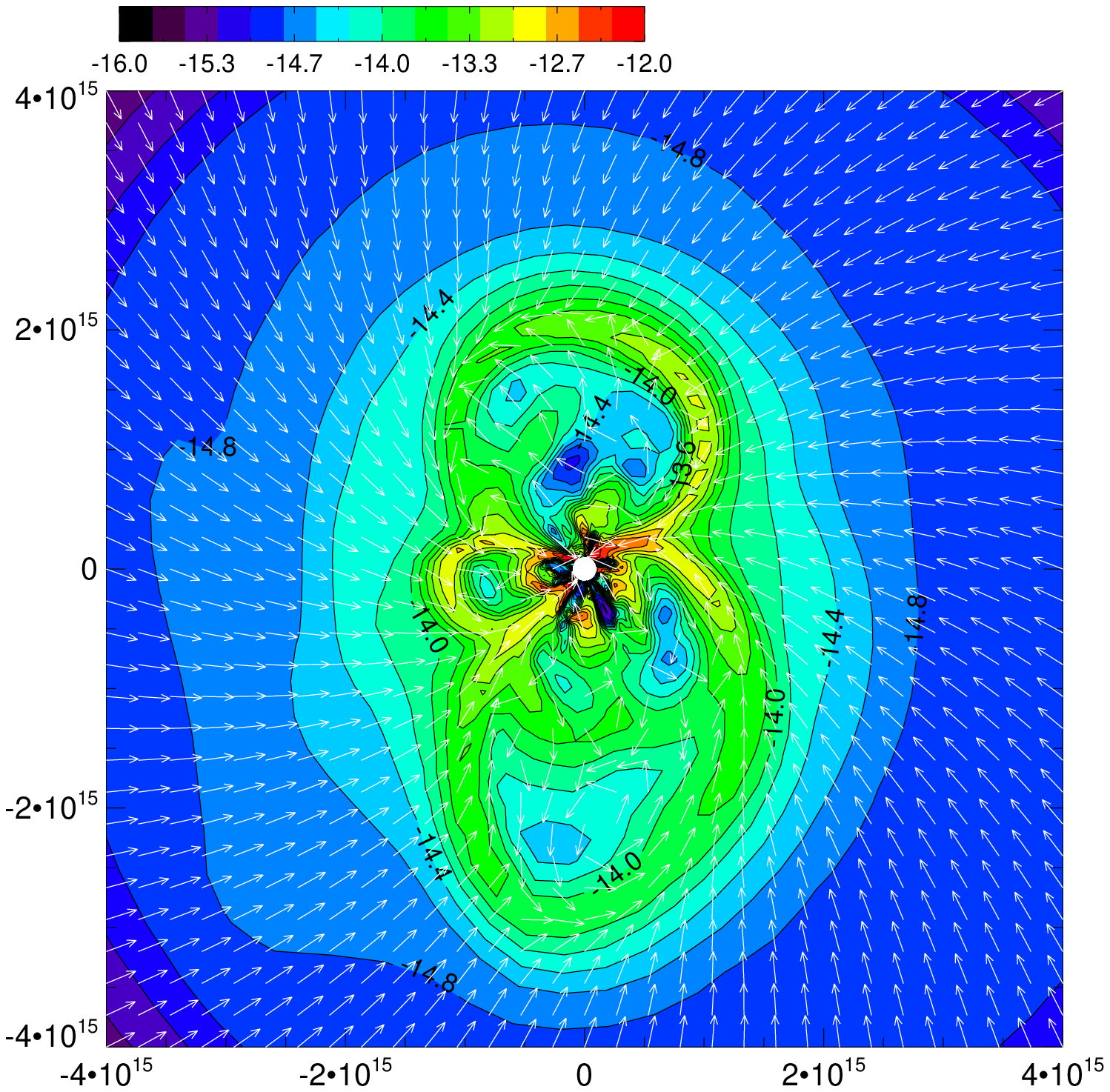}
\caption{Distribution of the logarithm of the mass density $\rho$ and
velocity field (unit vectors) on the equatorial plane, for Models
I (left panel, non-rotating) and J (right, rotating) that include both
ambipolar diffusion and (enhanced) Ohmic dissipation for magnetic
decoupling at small
radii. Again, both models show filamentary structures in the inner
protostellar accretion flow driven by magnetic interchange
instability.
}
\label{fig:Models_IJ}
\end{figure}

In Models I and J, we repeat Model G and H, but include ambipolar
diffusion with a cosmic ray ionization rate of $\zeta=9\times
10^{-17}\second^{-1}$ (see Table \ref{table:first}). These are the most comprehensive
of our simulations, because they include both ambipolar diffusion
that is important at relatively low densities and Ohmic dissipation
that is thought to play a crucial role in the magnetic decoupling
at high densities (\ct{Nakano+2002}). The results are illustrated in
Fig.\ \ref{fig:Models_IJ}. The left panel shows the non-rotating Model I
at a time $t=4.32\times 10^{12}\second$, when the central mass is
$0.162\msun$. The right panel displays the rotating Model J at
$t=4.67\times 10^{12}\second$, when the central mass is
$0.129\msun$. In both cases, the flow morphology is dominated
by expanding lobes along some azimuthal directions and infall along
others, broadly similar to the features present in all other models.
The similarity reinforces the notion that these are robust features
that are insensitive to the detailed treatment of the magnetic
decoupling, the nature of the microscopic magnetic diffusion
(ambipolar diffusion or Ohmic dissipation), or rotation.

\section{Discussion and Summary}
\label{discussion}

\subsection{Magnetic Domination of Inner Protostellar Accretion Flow}

A general result that we find is that the magnetic field dominates
the dynamics of the inner protostellar accretion flow out to hundreds
of AU (see, e.g.,
Fig.\ \ref{fig:ModelB_color}), even though its initial strength is relatively
moderate, corresponding to a dimensionless core mass-to-flux ratio
$\lambda \sim$ 3--4. This result is related to the magnetic flux 
problem, which lies at the heart of magnetized star formation.
As mentioned in \S \ref{intro}, if the magnetic flux
of a typical dense star-forming
core were to be carried into the central star, the stellar field strength
would be orders of magnitude higher than the observed values (see,
e.g., \S 4 of \ct{Nakano1984}). The vast majority of the core
magnetic field must be decoupled from the central mass.
What happens to the decoupled magnetic flux?

The decoupled flux can in principle be trapped by the protostellar
accretion flow through ram pressure. For a simple estimate, we note
that the magnetic flux associated with a stellar mass of $M_*$ is
given by $\Phi_* = 2\pi G^{1/2} M_*/\lambda$, where $\lambda$ is the
dimensionless mass-to-flux ratio of the star-forming core. If, after
decoupling from the stellar mass, this flux is confined within a
(cylindrical) radius of $R$, the field strength inside the radius
would be $B\approx \Phi_*/(\pi R^2)$, and the associated magnetic
pressure would be
\begin{equation}
P_B\approx\frac{GM_*^2}{2\pi \lambda^2 R^4}\ .
\label{MagneticPressure}
\end{equation}
This magnetic pressure is to be compared with the ram pressure of the
protostellar accretion flow at the same radius
\begin{equation}
P_R = \rho v_r^2 \approx
\frac{\sqrt{2}}{4\pi h}
\frac{G^{1/2} M_*^{1/2}{\dot M}}{R^{5/2}}\ ,
\label{RamPressure}
\end{equation}
where ${\dot M}$ is the rate of mass accretion (which occurs mostly
through a dense, flattened pseudodisk), and $h$ is the half-thickness
of the pseudodisk relative to the radius $R$. The infall speed
$v_r$ is assumed to be close to the free fall speed $v_{\rm{ff}}=(2
GM_*/R)^{1/2}$. The ratio of the magnetic to ram pressure is therefore
\begin{equation}
\xi \approx
\frac{\sqrt{2} h G^{1/2} M_*^{3/2}}{\lambda^2 {\dot M} R^{3/2}}
 = 1.14\times 10^2
\left(\frac{h}{0.1}\right)
\left(\frac{M_*}{0.5\msun}\right)^{3/2}
\left(\frac{4}{\lambda}\right)^2
\left(\frac{10^{-5}\msun\yr^{-1}}{\dot M}\right)
\left(\frac{10^{14}\cm}{R}\right)^{3/2}\ .
\label{PressureRatio}
\end{equation}
Note that the magnetic pressure increases with decreasing radius
faster than the ram pressure, indicating that it is more difficult
to confine the decoupled flux to a smaller radius. For example,
at the inner boundary of our simulation domain ($R = 10^{14}\cm$),
the decoupled flux would produce a magnetic pressure larger than
the ram pressure by two orders of magnitude for typical parameters;
it cannot be confined there. The same is even more true if the inner
edge of the simulation is chosen to be closer to the protostar.
The decoupled flux must therefore expand to a region well beyond
the inner boundary of our simulation. The characteristic size
of the region can be estimated by setting the ratio $\xi$ in
equation (\ref{PressureRatio}) to 1 (see also \ct{LiMcKee1996},
their equation 9):
\begin{equation}
R_B \approx
2.35\times 10^{15}
\left(\frac{h}{0.1}\right)^{2/3}
\left(\frac{M_*}{0.5\msun}\right)
\left(\frac{4}{\lambda}\right)^{4/3}
\left(\frac{10^{-5}\msun\yr^{-1}}{\dot M}\right)^{2/3} \ \ {\rm{cm}}\ .
\label{AlfvenRadius}
\end{equation}
At this radius, the ratio of the magnetic to thermal pressure would be
\begin{equation}
\xi = 1.41\times 10^2
\left(\frac{0.2\kms}{a}\right)^2
\left(\frac{0.1}{h}\right)^{2/3}
\left(\frac{\lambda}{4}\right)^{4/3}
\left(\frac{\dot M}{10^{-5}\msun\yr^{-1}}\right)^{2/3},
\label{PlasmaBeta}
\end{equation}
which is much greater than unity for typical parameters. We therefore
expect the decoupled magnetic flux to dominate the dynamics of
the inner protostellar accretion flow up to $R_B$, if it can be
transported outward to such a radius.

\subsection{Two Modes of Magnetic Flux Transport: Microscopic
  Diffusion vs Macroscopic Advection}
\label{modes}

How the magnetic flux is transported outward against the collapsing
inflow depends critically on the dimensionality assumed for the
problem. In a 2D (strictly axisymmetric) collapsing flow, the flux
can be transported outward only through microscopic non-ideal
MHD processes, such as ambipolar diffusion or Ohmic dissipation,
which allow the bulk material to cross the field lines. (We note that,
in 2D ideal MHD simulations, magnetic diffusivity and reconnection
of numerical origin can have a similar effect.) For realistic
levels of cloud core ionization, the microscopic magnetic diffusion
coefficient over most of the collapsing flow is rather small, however
(see, e.g., \ct{Li+2011}).
The magnetic flux dragged in by the collapsing flow can diffuse
outward only slowly. As a result, most of it is confined to a
small circumstellar region where the field strength is high and,
in the case of ambipolar diffusion, the diffusion rate is
enhanced by a large field gradient.
The situation is analogous to the energy transport by
radiative diffusion inside a star: for a small radiative diffusion
coefficient (or a large Rosseland mean opacity), a large temperature
gradient is required to transport a given energy flux. When the
temperature gradient in a region becomes too large, the matter would
turn convective, with energy advected outward by bulk fluid
motions. If the star were to be held strictly spherically symmetric,
this second mode of energy transport would be completely suppressed.

Similarly, we have demonstrated that, when the assumption of
axisymmetry is lifted, the magnetic flux in a protostellar
accretion flow can be transported outward by macroscopic
advection as well. We showed that initially
smooth axisymmetric protostellar accretion flows break up
spontaneously, with the more strongly magnetized regions expanding
away from the central gravitating object along some azimuthal
directions and the less magnetized region sinking toward it along
others. This simultaneous sinking and rising of material of
different degrees of magnetization is a classical
sign of the well-known magnetic buoyancy or interchange
instability (e.g., \ct{Parker1979}; \ct{Kaisig+1992};
\ct{StehleSpruit2001}; \ct{deGouveiaDalPino+2011}). The flow
pattern leads to an efficient outward transport of magnetic flux
relative to matter, even in a region where the microscopic diffusion
is absent (see \S \ref{filamentary}).

\subsection{Implications on Disk Formation}

Our results have implications on a problem of considerable current
interest: protostellar disk formation.
For the observationally inferred level of magnetization in dense cores,
disk formation is difficult in the strict ideal MHD limit, because a
magnetic split-monopole is expected to form, which can remove essentially
all of the angular momentum of the infalling material through magnetic
braking (see discussion in \S \ref{intro} and references therein).
Rotationally supported disks are formed in some ideal MHD simulations
(\ct{Machida+2011}), particularly when the rotation and magnetic
axes are misaligned (\ct{Joos+2012}) or in the presence of a strong
turbulence (\ct{Seifried+2012}, \ct{deGouveiaDalPino+2011}).
However, the expected magnetic split-monopole is not clear in these
calculations, which is a concern.

It was hoped that non-ideal MHD effects may weaken the magnetic braking
enough to enable disk formation. \citet{Machida+2007} and \citet{Dapp+2012}
showed that Ohmic dissipation can enable the formation of a
small (AU-scale) rotationally supported disk (RSD) in a region where
the column density is high enough to shield out the ionizing cosmic rays.
\citet{Krasnopolsky+2010} demonstrated that if the resistivity is
significantly enhanced, it is possible to form even large, 100-AU
sized RSDs. Such disks can also form in principle through spin-up
caused by the Hall effect, if the Hall coefficient is large enough
(\ct{Krasnopolsky+2011}). However, the microscopic values of the
resistivity and
Hall coefficient do not appear high enough for large RSDs to form
(\ct{Li+2011}). Furthermore, ambipolar diffusion, the most widely
studied non-ideal MHD effect in star formation, appears to make
disk formation more (rather than less) difficult (see discussion in
\S \ref{intro} and references therein). Classical non-ideal MHD
effects may not enable disk formation, at least under the assumption
of axisymmetry.

In the absence of axisymmetry, we find that the structure of the
protostellar accretion flow is modified considerably by a new ingredient:
interchange instability. This instability is expected to make disk
formation easier, because it enables the magnetic flux accumulated
near the protostar to be advected by the bulk fluid motions to a
larger distance, which lowers the field strength (see
Fig.\ \ref{fig:ModelB_lines}) and thus the magnetic braking efficiency.
However, we have carried out a number of simulations that include
rotation (see Table \ref{table:first}), and found no evidence for the formation of a
rotationally supported disk, even at relatively late times. This
is in agreement with the AMR MHD simulations of \citet{Zhao+2011},
who noted that the strong magnetic field in the low-density
expanding lobes prevents the rotating infalling material from making
a full orbit around the center (see their Fig.\ 6 and the last panel
in our Figs.\ \ref{fig:Models_CDEF} and \ref{fig:Models_IJ}).
Our calculations indicate that the interchange instability in 3D
may not weaken the magnetic braking enough to enable disk
formation, although this issue deserves a closer examination.

Another implication is that, in the presence of interchange
instability, ambipolar diffusion becomes less important in
transporting magnetic flux. The reason is that the
instability allows the magnetic flux to be advected outward,
reducing the gradient in the field that is needed to drive
the ambipolar diffusion. This result provides some
justification for the 3D ideal MHD calculations of protostellar
mass accretion, such as those of \citet{Zhao+2011}, that ignore
ambipolar diffusion, as long as the magnetic decoupling at high
densities is accounted for.

Finally, we mention in passing that the filamentary protostellar
accretion flow structured by the interchange instability occurs
on the scales of
order $10^2\au$ or larger, which can in principle be probed with
the Atacama Large Millimeter/submillimeter Array in nearby star
forming clouds. Observational studies of such a region may be
important for understanding magnetized accretion onto not only
protostars, but also other astrophysical objects, such as active
galactic nuclei and the black hole at the Galactic center
(e.g., \ct{IgumenshchevNarayan2002}; \ct{Pang+2011}; \ct{McKinney+2012}).

\subsection{Summary}

We have carried out three dimensional simulations of
the collapse of magnetized dense cores including three nonideal MHD 
processes: ambipolar diffusion, Ohmic dissipation, and decoupling 
at the inner boundary. 
Our main result is that the inner protostellar
accretion flow is driven unstable by the magnetic flux decoupled
from the matter that enters the central object. The instability
is of the interchange type. When it is fully developed, the flow
structure becomes highly filamentary, as a result of the interplay
between gravity driven infall and magnetically driven expansion.
We showed, in particular, that the magnetically-dominated structure
inside the ambipolar diffusion-induced hydromagnetic shock found
in previous axisymmetric studies is unstable in 3D, as it has been
anticipated for some time. Without the restriction of axisymmetry,
the redistributed magnetic flux can be transported outward
advectively, through the bulk motions of low-density expanding
regions. This new channel of efficient flux transport renders
the microscopic processes, such as ambipolar diffusion, less directly
important in redistributing magnetic flux in the protostellar
accretion flow. It also lowers the magnetic field strength close
to the protostar, which could in principle make the magnetic
braking less efficient and the formation of a rotationally
supported disk easier. However, we find no evidence for disk
formation in any of our rotating collapse simulations. How a
rotationally supported disk forms in a largely magnetically
dominated, filamentary protostellar accretion flow is an
outstanding unsolved problem.

\acknowledgments
The work was supported in part by NASA through NNX10AH30G and by the
Theoretical Institute for Advanced Research in Astrophysics (TIARA)
under the CHARMS initiative and the National Science Council of Taiwan
through grant NSC97-2112-M-001-018-MY3.

\end{document}